\documentclass[a4paper,11pt]{article}
\pdfoutput=1
\usepackage{latexsym}
\usepackage{amsmath, amsthm, amssymb, bbm}
\usepackage[mathscr]{eucal}
\usepackage{tikz}
\usepackage{cite}
\usepackage{hyperref}
\usepackage{graphicx}
\usepackage[margin=3cm]{geometry}

\theoremstyle{plain}

\theoremstyle{definition}

\theoremstyle{remark}

\newcommand{\rhs}{r.h.s.\ }
\newcommand{\lhs}{l.h.s.\ }
\newcommand{\wrt}{w.r.t.\ }
\newcommand{\cf}{cf.\ }

\newcommand{\bra}[1]{\langle #1 \rvert}
\newcommand{\ket}[1]{\lvert #1 \rangle}

\newcommand{\VEV}[1]{\langle #1 \rangle}

\newcommand{\ud}{\mathrm{d}}
\newcommand{\del}{\partial}

\newcommand{\betrag}[1]{{\lvert #1 \rvert}}

\newcommand{\R}{\mathbb{R}}

\newcommand{\order}{\mathcal{O}}

\newcommand{\1}{\mathbbm{1}}

\newcommand{\sS}{\mathcal{S}}

\newcommand{\cR}{\mathcal{R}}

\newcommand{\eps}{\varepsilon}

\newcommand{\N}{\mathbb{N}}

\newcommand{\bk}{{\mathrm{bk}}}
\newcommand{\bd}{{\mathrm{bd}}}

\newcommand{\vp}{{\varphi}}

\newcommand{\nn}{\nonumber}
\newcommand{\beq}{\begin{equation}}
\newcommand{\eeq}{\end{equation}}

\newcommand{\defeq}{\mathrel{:=}}

\begin{document}

\title{The semi-classical energy of the open Nambu-Goto string}
\author{Jochen Zahn \\ Institut f\"ur Theoretische Physik, Universit\"at Leipzig \\ Br\"uderstr.~16, 04103 Leipzig, Germany \\ jochen.zahn@itp.uni-leipzig.de}

\date{\today}

\maketitle

\begin{abstract}
We compute semi-classical corrections to the energy of rotating open Nambu-Goto strings with and without masses at the ends, using methods from quantum field theory on curved space-times.
\end{abstract}

\section{Introduction}

For several reasons, the Nambu-Goto string is an interesting model: It exhibits diffeomorphism invariance, making it a toy model for (quantum) gravity. It also provided motivation for the Polyakov string, which led to string theory as a candidate for a fundamental theory. Furthermore, it constitutes a phenomenological model for QCD vortex lines connecting quarks, i.e., for the description of hadrons. In this context, the generalization to masses at the endpoints \cite{ChodosThorn74} is particularly interesting \cite{SonnenscheinWeissman14}.

It is well-known \cite{Rebbi74, Scherk75} that in the covariant quantization of the open massless Nambu-Goto string, the intercept $a$ is a free parameter, only constrained by the fact that the theory is consistent only for $a \leq 1$ and $D \leq 25$ or $a = 1$ and $D = 26$. Furthermore, the ground state energies $E_{\ell_{1, 2}}$ for a given angular momentum $\ell_{1, 2} > 0$, say in the $1-2$ plane, lie on the Regge trajectory
\beq
\label{eq:ReggeTrajectory}
 E_{\ell_{1, 2}}^2 = 2 \pi \gamma (\ell_{1, 2} - a),
\eeq
with $\gamma$ the string tension.

In both light cone gauge quantization and the Polyakov string, the allowed range for $a$ and $D$ shrinks to the critical case $a =1$ and $D=26$. However, for the application as a phenomenological model in QCD, it is certainly desirable to compute $a$ for $D< 26$, in particular $D = 4$, at least to leading order. There are two different frameworks for the treatment of the bosonic string at non-critical dimension. The approach taken here is based on the finding that, as an effective field theory, in the sense of perturbation theory around arbitrary non-degenerate classical solutions, the Nambu-Goto string is anomaly-free in any target space dimension \cite{BRZ}.
Perturbation theory is there based on splitting the embedding $X: \Sigma \to \R^D$ into a classical solution $\bar X$ and a perturbation $\vp$, i.e.,
\beq
\label{eq:X_lambda_phi}
 X = \bar X + \gamma^{-\frac{1}{2}} \vp,
\eeq 
and to quantize the perturbation $\vp$.

The other approach to effective string theory in non-critical dimensions is defined by the Polchinski-Strominger (PS) action \cite{PolchinskiStrominger}. It is derived by 
fixing the parametrization to conformal gauge and introducing singular supplementary terms in order to preserve the conformal symmetry at the quantum level. Concretely, it is the full embedding $X: \Sigma \to \R^D$, \cf \eqref{eq:X_lambda_phi}, which is gauge fixed. In particular, this implies constraints on the parametrization of the classical embedding $\bar X$.

Conceptually, this is quite different from our approach to effective string theory \cite{BRZ}: There, gauge conditions are only imposed on the fluctuations $\vp$, the parametrization of the classical solution $\bar X$ being arbitrary. This corresponds to the standard treatment of Yang-Mills theories in background fields or of perturbative quantum gravity.\footnote{We refer to \cite{DubovskyFlaugerGorbenko, AharonyKomargodski, HellermanShunsukeMaltzSwanson} for further discussions of the relations between different effective string theories.}

For the PS action, the intercept
\beq
\label{eq:a1}
a = 1
\eeq
was obtained in \cite{HellermanSwanson} for rotating open strings, independently of the dimension $D$. The corrections due to the supplementary PS term were evaluated in a classical rotating solution. Boundary divergences appearing in this calculation were removed by a boundary counterterm. To this, the Casimir contribution $a_{\mathrm{Cas}} = \frac{D-2}{24}$ was added. The latter is obtained in the ground state (not on a rotating background). Therefore, we think that the result is not obviously interpretable as a semi-classical value, in the sense that it is not obtained by perturbation theory around classical solutions. Thus, it seems desirable to check the result by a direct semi-classical calculation, and we can indeed confirm \eqref{eq:a1}. Apart from that, our approach has further benefits:
\begin{itemize}
\item We obtain information about the spectrum of physical excitations.
\item The issue of renormalization ambiguities and boundary counterterms may be more transparent in our approach.
\item Our approach allows to also treat open strings with masses at the endpoints. In particular, one may confirm that in the massless limit, one obtains the result for the massless string.\footnote{This is not obvious, since, as noted below, the boundary conditions for the massive string do not converge to the open string boundary conditions in the massless limit.}
\item It provides an analytically tractable toy model for locally covariant renormalization on curved space-times \cite{HollandsWaldWick}.
\end{itemize}

Let us briefly describe our approach.
Our starting point are classical rotating string solutions for the Nambu-Goto string. We then quantize the perturbations to these solutions at second order in the perturbation, obtaining a free quantum field living on the world-sheet. This is a curved manifold, and the equations of motion for the fluctuations only depend on the world-sheet geometric data, i.e., the induced metric and the second fundamental form. Hence, it seems natural, in line with the framework of \cite{BRZ}, to use methods from quantum field theory on curved space-time \cite{HollandsWaldWick, HollandsWaldReview} for the renormalization of the free world-sheet Hamiltonian $H^0$. The crucial requirements are that the renormalization is performed in a local and covariant way, and that the renormalization conditions are fixed only once. The latter means that they are ``the same'' on all classical solutions for the same bare parameters (in the present case, the only bare parameters are the string tension and possibly the masses at the ends). We find that there are only two renormalization ambiguities in $H^0$, amounting to geodesic curvature counterterms on the two boundaries. For identical masses at the boundaries, this reduces to a single ambiguity, which amounts to an Einstein-Hilbert counterterm. Furthermore, the energy density is locally finite but diverges in a non-integrable fashion at the boundaries. In line with the usual treatment of such divergences \cite{DowkerKennedy78}, we regularize them by introducing geodesic curvature counterterms at the boundaries. The correspondence between the world-sheet Hamiltonian and the target space energy then gives corrections to the classical Regge trajectories.

Let us analyze this in a bit more detail. To simplify the discussion, we here restrict to the massless string. The modifications due to masses at the endpoints are discussed in Section~\ref{sec:ClassicalSolutions}. The classical target space energy and angular momentum for the string rotating in the $1 - 2$ plane are
\begin{align}
\label{eq:E_class_massless}
 \bar E & = \gamma \pi R, &
 \bar L_{1, 2} & = \tfrac{1}{2} \gamma \pi R^2,
\end{align}
with $2 R$ the string length in target space. In the parametrization that we are using, the world-sheet time $\tau$ is dimensionless, and so should be the world-sheet Hamiltonian $H$, which generates translations in $\tau$. Its free part $H^0$ does not contain any further parameters, the string tension $\gamma$ appearing in inverse powers in the interaction terms. By dimensional analysis, we must thus have
\[
 H = H^0 + \order(R^{-1} \gamma^{-\frac{1}{2}}),
\]
with $H^0$ independent of $R$ and $\gamma$.\footnote{In principle, also a term $\log \Lambda R$, with $\Lambda$ a renormalization scale, might be induced by renormalization. This would imply that the intercept is ambiguous. However, we find that such a term is not present.} In our parametrization, the relation between the world-sheet Hamiltonian $H$, the quantum correction $E^q$ to the target space energy $E$, and the quantum correction $L^q_{1,2}$ to the angular momentum $L_{1,2}$ is 
\beq
 \label{eq:H_E_L}
 E^q = \tfrac{1}{R} (H + L^q_{1,2}),
\eeq
leading to
\begin{align*}
 E^2 & = (\bar E + E^q)^2  \nn \\
 & = \gamma^2 \pi^2 R^2 + 2 \gamma \pi (H + L^q_{1,2}) + \order(R^{-2}) \nn \\
% \label{eq:E_squared}
 & = 2 \gamma \pi (L_{1, 2}+ H^0) + \order(\bar L_{1, 2}^{-\frac{1}{2}}).
\end{align*}
By comparison with \eqref{eq:ReggeTrajectory}, one can directly read off the intercept $a$ from the expectation value of $H^0$, i.e.,
\beq
\label{eq:a_H0}
 a = - \VEV{H^0}.
\eeq
As already stated, our method yields the intercept \eqref{eq:a1}, independently of the dimension, consistent with the result obtained using the PS action \cite{HellermanSwanson}.

Let us comment on the relation to other semi-classical calculations of the intercept. In \cite{Nesterenko91}, the non-relativistic limit of the rotating string with masses at the ends was considered.
The calculation of the energy proceeds via the series of eigenfrequencies. Now there are many different ways to regularize such a series, so without any physical input, one can get an arbitrary dependence of the energy on the angular momentum. This is exemplified by considering two, mathematically well-motivated, schemes, that lead to qualitatively different results. This constitutes a good example for the need for a physically motivated renormalization scheme in order to obtain unambiguous results. We think that our local renormalization scheme fulfills this criterion.

In \cite{BakerSteinke02}, building on results in \cite{BakerSteinke01}, the full relativistic problem was considered. This work is closest in spirit to our calculation, so we discuss the differences in some detail. The quantization of the fluctuations around the rotating string solution with masses at the ends there led to the intercept
\beq
\label{eq:aClassical}
 a = \frac{D-2}{24},
\eeq
which would be consistent with the above mentioned results for $D=26$. However, some comments are in order. First, for the fluctuations, Dirichlet boundary conditions are imposed. These are not the ones that one obtains with masses at the ends \cite{RotatingString}. Second, the renormalization, in particular of  the logarithmic divergences, is not manifestly local on the world-sheet.\footnote{One mistake in the calculation was already pointed out in \cite{HellermanSwanson}. But there is further mistake in the treatment of the logarithmic divergences, \cf Footnote~\ref{ft:BS} below.} 
Third, for the corrections to the energy as a function of the classical angular momentum $\bar L$,
\beq
\label{eq:aBS}
 \bar a = \frac{1}{2} + \frac{D-2}{24} 
\eeq
is obtained. The result \eqref{eq:aClassical} is then gotten upon the replacement 
\beq
\label{eq:Langer}
 \bar L = \ell + \tfrac{1}{2}.
\eeq
While this so-called Langer modification is well known in semi-classical calculations, it applies to quantum mechanical problems in three spatial dimensions if no fluctuations perpendicular to the plane of rotation are allowed.
All these criteria are not fulfilled in the setting of \cite{BakerSteinke01, BakerSteinke02}, so the substitution \eqref{eq:Langer} does not seem to be justified.\footnote{In Section~\ref{sec:Hydrogen}, we will see that if, in the quantum mechanical context, fluctuations perpendicular to the plane of rotation are allowed, then the correct result is obtained without the Langer modification.} Finally, let us remark that the additional term $\frac{1}{2}$ in \eqref{eq:aBS} is due to the fact that a mode with frequency equal to the rotation frequency of the classical solution is absent from the spectrum.

In \cite{ZayasSonnenscheinVaman}, the fluctuations around solutions to the massless Nambu-Goto string were quantized. The calculation of the intercept then proceeded by $\zeta$ function regularization of the series of eigenmodes,\footnote{The problem with such calculations was already discussed above.} leading to \eqref{eq:aBS}. As before, the reason is the absence of a certain mode.  A similar calculation is also performed for the Polyakov action, leading to the intercept \eqref{eq:aClassical}.

Hence, both in \cite{BakerSteinke02} and \cite{ZayasSonnenscheinVaman}, a certain mode, that one might naively expect to be present, is absent. In our terminology, introduced below, this is the planar $n=1$ mode, and it is also absent in our approach to the massless string. However, for the massive string, the mode is present and can be interpreted as a Nambu-Goldstone mode for the broken translation invariance in the plane of rotation \cite{RotatingString}.
We will show that there is a corresponding linearly growing mode and that the two modes form a canonical pair, i.e., there is indeed no ground state energy corresponding to this mode. Furthermore, we point out that this mode is also absent, more precisely represented by a null state, in the covariantly quantized open string for the intercept \eqref{eq:a1}, \cf Section~\ref{sec:Degeneracy}.

The article is structured as follows: In the next section, we discuss, as a motivating example for our semi-classical calculation, the hydrogen atom. The aim is to introduce some of the terminology used later on, such as the distinction of the world-sheet Hamiltonian and the target space energy. Furthermore, it shows that some, at first sight disturbing, features we will encounter in the discussion of the Nambu-Goto string, are in fact generic for a semi-classical analysis. In Section~\ref{sec:ClassicalSolutions}, we discuss the classical rotating string solutions for the case of masses at the endpoints. In Section~\ref{sec:Quantization}, the fluctuations of classical rotating string solutions and their canonical quantization are discussed. In Section~\ref{sec:Energy}, the relation between the world-sheet Hamiltonian and the target space energy is discussed. In Section~\ref{sec:Massless} the locally covariant renormalization of the world-sheet Hamiltonian is explained and performed in the massless case, yielding the intercept~\eqref{eq:a1}. In Section~\ref{sec:Massive}, the massive string is treated. Section~\ref{sec:Degeneracy} is devoted to the comparison of the excitation spectra of the semi-classical string and that of the covariantly quantized string.
An appendix contains some calculations that were omitted in the main part.

\section{A motivating example: The hydrogen atom}
\label{sec:Hydrogen}

As a motivating example for our semi-classical approach, we consider the hydrogen atom in a semi-classical approximation. We do not do this in the most straightforward way, but rather in a fashion that is very close to our treatment of Nambu-Goto string. In particular, we introduce a \emph{parametrization time} $\tau$ and consider trajectories $t(\tau), \vec x(\tau)$ in the \emph{target space-time} $\R \times \R^3$. The Lagrangian for these is given by
\[
 \mathcal{L} = \frac{1}{2} \frac{\betrag{\dot{\vec x}}^2}{\dot t} + \frac{1}{\betrag{\vec x}} \dot t,
\]
where the dot denotes derivatives \wrt $\tau$. The \emph{energy} and \emph{angular momentum} in the $1-2$ plane is given by
\begin{align*}
 E & = - \frac{\del \mathcal{L}}{\del \dot t} =  \frac{1}{2} \frac{\betrag{\dot{\vec x}}^2}{\dot t^2} + \frac{1}{\sqrt{\rho^2 + z^2}}, \\
 L & = \frac{\del \mathcal{L}}{\del \dot \phi} = \rho^2 \frac{\dot \phi}{\dot t},
\end{align*}
where we switched to cylindrical coordinates $(\rho, \phi, z)$.

There are circular orbits extremizing the above action. The aim is to quantize the perturbations around these. We parametrize them as
\begin{align*}
 \rho(\tau) & = R \left( 1 + \gamma^{-\frac{1}{2}} r(\tau) \right), \\
 \phi(\tau) & = \tau + \gamma^{-\frac{1}{2}} \vp(\tau), \\
 z(\tau) & = R \gamma^{-\frac{1}{2}} \xi(\tau), \\
 t(\tau) & = R^{\frac{3}{2}} \left( \tau + \gamma^{-\frac{1}{2}} \vartheta(\tau) \right)
\end{align*}
with
\[
 \gamma = R^{\frac{1}{2}}.
\]
Expanding the Lagrangian up to second order in the perturbations $(r, \vp, \xi, \vartheta)$, we obtain an irrelevant constant, a first order term which is a total derivative, and the second order term
\beq
\label{eq:HydrogenH0_0}
 \mathcal{L}^0 = \frac{1}{2} \left( \dot r^2 + (\dot \vp - \dot \vartheta)^2 + \dot \xi^2 - \xi^2 + 3 r^2 + 4 r (\dot \vp - \dot \vartheta) \right),
\eeq
henceforth also called the free part. We see that the combination $\vp + \vartheta$ does not appear in the Lagrangian, so we may consistently set it to zero, i.e., perform the gauge fixing $\vartheta = - \vp$, and redefine
\begin{align}
%\label{eq:Hydrogen_phi}
 \phi(\tau) & = \tau + \tfrac{1}{2} \gamma^{-\frac{1}{2}} \vp(\tau), \nn \\
\label{eq:Hydrogen_t}
 t(\tau) & = R^{\frac{3}{2}} \left( \tau - \tfrac{1}{2} \gamma^{-\frac{1}{2}} \vp(\tau) \right),
\end{align}
yielding the free Lagrangian
\beq
\label{eq:HydrogenLag}
 \mathcal{L}^0 = \frac{1}{2} \left( \dot r^2 + \dot \vp^2 + \dot \xi^2 - \xi^2 + 3 r^2 + 4 r \dot \vp \right).
\eeq
The free Hamiltonian corresponding to this action is
\beq
\label{eq:HydrogenH0}
 H^0 = \frac{1}{2} \left( \dot r^2 + \dot \vp^2 + \dot \xi^2 + \xi^2 - 3 r^2 \right).
\eeq
Each supplementary order in the perturbations $(r, \vp, \xi)$ is suppressed by a factor of $\gamma^{-\frac{1}{2}}$, which we use as the formal expansion parameter in our perturbative treatment.

For the expansion of the energy and the angular momentum in the perturbations $(r, \vp, \xi)$, we obtain
\begin{align}
\label{eq:HydrogenE}
 E & = \frac{1}{R} \left[ -\frac{1}{2} + \gamma^{-\frac{1}{2}} ( 2 r + \dot \vp ) + \gamma^{-1} \frac{1}{2} \left( \dot r^2 + 2 \dot \vp^2 + \dot \xi^2 + 4 r \dot \vp - r^2 + \xi^2 \right) \right] + \order(R^{-\frac{7}{4}}) , \\
\label{eq:HydrogenL}
 L & = \gamma \left[ 1 + \gamma^{-\frac{1}{2}} ( 2 r + \dot \vp ) + \gamma^{-1} \frac{1}{2} \left( \dot \vp^2 + 4 r \dot \vp + 2 r \right) \right] + \order(R^{-\frac{1}{4}}).
\end{align}
We separate $E$ and $L$ into the classical parts $\bar E$ and $\bar L$, which are independent of the perturbations $(r, \vp, \xi)$, and the remainder $E^q$ and $L^q$, i.e.,
\begin{align*}
 E & = \bar E + E^q, &
 L & = \bar L + L^q.
\end{align*}
Obviously, we have
\beq
\label{eq:E_L_Hydrogen}
 \bar E = - \frac{1}{2} \frac{1}{\bar L^2},
\eeq
the classical relation between energy and angular momenta for circular orbits.

In the theory where the perturbations are quantized, $E^q$ and $L^q$ should generate target space time translations and rotations, while the Hamiltonian $H$ should generate parametrization time translations. From the relation \eqref{eq:Hydrogen_t}, it follows that, up to a scale, a target time translation corresponds to a parametrization time translation. However, the classical solution also rotates, so the correct relation between the energy correction $E^q$ and the Hamiltonian $H$ is
\beq
\label{eq:Relation_H_E_Hydrogen}
 R^{\frac{3}{2}} E^q = H + L^q.
\eeq
This is clearly the analog of \eqref{eq:H_E_L}. That this relation is correct up to second order in the perturbation $(r, \vp, \xi)$ can easily be checked from \eqref{eq:HydrogenH0}, \eqref{eq:HydrogenE}, \eqref{eq:HydrogenL}. Furthermore, the parametrization time translation $\tau \mapsto \tau + 2 \pi$ corresponds to the target space time translation $t \mapsto t + 2 \pi R^{\frac{3}{2}}$. Comparing with \eqref{eq:Relation_H_E_Hydrogen}, we find that the spectrum of $L^q$ should be the integers, as expected for an angular momentum operator.

Let us expand $E^q = E^q_1 + E^q_2 + \dots$ in powers of $\gamma^{-\frac{1}{2}}$ and likewise for $L^q$. Assume that we may choose an eigenstate of $E^q_1$ (and hence also $L^q_1$) of eigenvalue $0$. This will be justified below. In such a state, we have
\begin{align}
 (- E)^{-\frac{1}{2}} & = (- \bar E)^{-\frac{1}{2}} + \tfrac{1}{2} (- \bar E)^{-\frac{3}{2}} E^q_2 + \order(R^{-\frac{1}{4}}) \nn \\
 & = 2^{\frac{1}{2}} \left( \bar L + R^{\frac{3}{2}} E^q_2 \right) + \order(R^{-\frac{1}{4}}) \nn \\
 \label{eq:Regge_Hydrogen}
 & = 2^{\frac{1}{2}} \left( L + H^0 \right) + \order(R^{-\frac{1}{4}}).
\end{align}
It follows that the semi-classical correction to the classical relation \eqref{eq:E_L_Hydrogen} can be computed by finding the ground state energy in the free Hamiltonian $H^0$.

Let us thus quantize the free theory defined by the Lagrangian \eqref{eq:HydrogenLag}. Setting $f = (r, \vp, \xi)$, the free equations of motion can be written as
\beq
\label{eq:eom_Hydrogen}
 \ddot f + A \dot f + B f = 0,
\eeq
with
\begin{align*}
 A & = \begin{pmatrix} 0 & - 2 & 0 \\ 2 & 0 & 0 \\ 0 & 0 & 0 \end{pmatrix}, &
 B & = \begin{pmatrix} - 3 & 0 & 0 \\ 0 & 0 & 0 \\ 0 & 0 & 1 \end{pmatrix}.
\end{align*}
Because of the term with a single time derivative in \eqref{eq:HydrogenLag}, canonical quantization has to be performed with some care. The symplectic form can be written as
\[
 \sigma((f, \dot f), (\tilde f, \dot{\tilde f})) = \begin{pmatrix} f & \dot f \end{pmatrix} 
 \begin{pmatrix}
 A & \1_3 \\
 - \1_3 & 0_3 
\end{pmatrix}
 \begin{pmatrix} \tilde f \\ \dot{\tilde f} \end{pmatrix}.
\]
Looking for mode solutions of the form
\[
 f_i(t) = f_i e^{- i \omega_i t},
\]
we find
\begin{align*}
 \omega_1 & = 0 &  f_1 & = (0,1,0), \\
 \omega_2 & = 1 & f_2 & = 2^{-\frac{1}{2}} (1,-2 i,0), \\
 \omega_3 & = 1 & f_3 & = 2^{-\frac{1}{2}} (0,0,1).
\end{align*}
Note that we already symplectically normalized the modes $f_2, f_3$, according to
\[
 \sigma(\bar f_i, f_j) = - i \delta_{i j}.
\]
This is of course not possible for the zero mode $f_1$. It is accompanied by a linearly growing mode, so that
\begin{align}
\label{eq:AngleL_Hydrogen}
 f_\theta & = \sqrt{3} (0, 1, 0), &
 f_\lambda & = \sqrt{3} (\tfrac{2}{3}, - t, 0).
\end{align}
form a symplectic pair, i.e.,
\beq
\label{eq:AngleL_Normalization_Hydrogen}
 \sigma(f_\theta, f_\lambda) = 1.
\eeq

We now write the general solution of the equation of motion \eqref{eq:eom_Hydrogen} as a linear combination of these modes, i.e.,
\[
 f = \left[ a_2 f_2 + a_3 f_3 + \text{h.c.} \right] + \theta f_\theta + \lambda f_\lambda.
\]
For the expansion of $E^q$ and $L^q$ in the perturbation we obtain
\begin{align*}
 E^q & = - 3^{-\frac{1}{2}} R^{-1} \gamma^{-\frac{1}{2}} \lambda + \order(R^{-\frac{3}{2}}), \\
 L^q & = - 3^{-\frac{1}{2}} R^{\frac{1}{2}} \gamma^{-\frac{1}{2}} \lambda + \order(R^0).
\end{align*}
We thus see that we should interpret $\lambda$ as the leading contribution to (a multiple of) the angular momentum operator $L^q$. Recalling that $L^q$ should have the integers as the spectrum, we conclude that $\lambda$ should be quantized not as a momentum operator, as suggested by \eqref{eq:AngleL_Normalization_Hydrogen}, but as an angular momentum operator, i.e., as a multiple of $- i \del_\phi$ on $L^2(S^1)$. In particular, $\hat \lambda$ should have an eigenvalue $0$. Restricting to the corresponding eigenstate amounts to fixing the angular momentum to its classical value up to corrections of $\order(R^0)$.

In terms of the coefficients $a_i$, $\theta$, $\lambda$, the free Hamiltonian reads
\beq
\label{eq:H0_Hydrogen}
 H^0 = a_2 \bar a_2 + a_3 \bar a_3 - \tfrac{1}{2} \lambda^2.
\eeq
The sign of the last term on the \rhs is due to the unusual sign of the linearly growing term in \eqref{eq:AngleL_Hydrogen}. Being in the $\hat \lambda$ eigenstate of eigenvalue $0$, we may ignore this term. The first two terms on the \rhs constitute a two-dimensional harmonic oscillator with frequency $\omega = 1$. Hence, we find that the $k$th excited state is $k+1$ times degenerate with eigenvalue $k+1$. By \eqref{eq:Regge_Hydrogen}, we thus find
\[
 E = - \frac{1}{2 (m + k + 1)^2} + \order(m^{- \frac{5}{2}})
\]
for the energy of the $k$th excited state with magnetic quantum number $m > 0$, i.e., the correct result. In particular, we see that the Langer modification \eqref{eq:Langer} yields the wrong result if perturbations perpendicular to the plane of rotation are allowed.

\section{The classical rotating string}
\label{sec:ClassicalSolutions}

The action for the Nambu-Goto string with masses at the ends \cite{ChodosThorn74} is given by
\beq
\label{eq:Action}
 \mathcal{S} = - \gamma \int_\Sigma \sqrt{\betrag{g}} - \sum_{c \in \pm} m_c \int_{\del_c \Sigma} \sqrt{\betrag{h}},
\eeq
where $\Sigma$ is the world-sheet, $\del_\pm \Sigma$ its two boundary components, $\gamma$ is the string tension, $m_\pm$ the masses at the two boundaries. Furthermore, $g$ is the induced metric in the bulk and $h$ the induced metric on the boundary. We work in signature $(-, +)$.

Following \cite{RotatingString}, it is convenient to parametrize the rotating string solution as
\begin{equation}
\label{eq:X}
 \bar X(\tau, \sigma) = R (\tau, \cos \tau \sin \sigma, \sin \tau \sin \sigma, 0),
\end{equation}
where $\sigma \in [-S_-, S_+]$, $S_\pm < \pi/2$. For simplicity, we here assumed that the target space-time is four dimensional. Adding further dimensions (or deleting one) is straightforward. \eqref{eq:X} is a solution to the above action, provided that
\beq
\label{eq:S_Condition}
 \frac{\gamma R}{m_\pm} = \frac{\tan S_\pm}{\cos S_\pm}.
\eeq
The induced metric on the world-sheet and on the boundary, in the coordinates introduced above, is
\begin{align}
\label{eq:metric}
 g_{\mu \nu} & = R^2 \cos^2 \sigma \eta_{\mu \nu}, &
 h & = - R^2 \cos^2 \sigma.
\end{align}
The bulk metric has scalar curvature
\[
%\label{eq:scalarCurvature}
 \cR = \frac{2}{R^2 \cos^4 \sigma}
\]
and the boundary component $c$ the geodesic curvature
\beq
\label{eq:GeodesicCurvature}
 \kappa_c = - \frac{\tan S_c}{R \cos S_c}.
\eeq

The (angular) momenta corresponding to the action~\eqref{eq:Action} are given by
\begin{align}
\label{eq:Momentum}
 P^i & = \int \frac{\delta \mathcal{S}}{\delta \del_0 X_i} \ud \sigma \\
  & = - \gamma \int_{-S_-}^{S_+} \sqrt{g} g^{0 \nu} \del_\nu X^i \ud \sigma + \sum_c m_c \betrag{h}^{-\frac{1}{2}} \del_0 X^i |_c, \nn \\
\label{eq:AngularMomentum}
 L_{ij} & = \int \frac{\delta \mathcal{S}}{\delta \del_0 X^j} X_i \ud \sigma - i \leftrightarrow j \\
 & = \gamma \int_{-S_-}^{S_+} \sqrt{g} g^{0 \nu} X_j \del_\nu X_i \ud \sigma - \sum_c m_c \betrag{h}^{-\frac{1}{2}} X_j \del_0 X_{i} |_c - i \leftrightarrow j . \nn
\end{align}
Here $\cdot |_c$ denotes the evaluation at $\sigma = c S_c$. The target space energy is given by $E = P^0$.

For the energy $\bar E$ and the angular momentum $\bar L = \bar L_{1,2}$ of the solution \eqref{eq:X}, one finds
\begin{align}
\label{eq:E_class}
 \bar E & = \sum_{c \in \pm} \left[ \gamma R S_c + \frac{m_c}{\cos S_c} \right] = \gamma R \sum_{c \in \pm} \left[ S_c + \frac{1}{\tan S_c} \right], \\
\label{eq:J_class}
 \bar L & = \sum_{c \in \pm} \left[ \frac{\gamma R^2}{2} \left( S_c - \frac{\sin 2 S_c}{2} \right) + m_c R \frac{\sin^2 S_c}{\cos S_c} \right]  = \frac{\gamma R^2}{2} \sum_{c \in \pm} \left[ S_c - \frac{\sin 2 S_c}{2} + \frac{\sin^2 S_c}{\tan S_c} \right].
\end{align}
In the massless limit ($m_\pm \to 0$ with $R, \gamma$ fixed) this reduces to \eqref{eq:E_class_massless}.
In particular, one finds the famous Regge trajectory
\[
%\label{eq:MasslessReggeTrajectory}
 \bar E^2 = 2 \pi \gamma \bar L.
\]
The Regge intercept $a$ is defined as the shift of the trajectory,
\beq
\label{eq:MasslessReggeTrajectoryShifted}
  E^2 = 2 \pi \gamma (L - a),
\eeq
possibly up to correction of $\order(L^{-1})$ (which are not present in the covariant quantization scheme).

To discuss the massive case, let us denote the two components of the energy and the angular momentum in \eqref{eq:E_class}, \eqref{eq:J_class} by $\bar E_\pm$ and $\bar L_\pm$.
For large $R$, we have
\begin{align}
\label{eq:E_c_Expansion}
 \bar E_c & = \frac{\pi \gamma}{2} R + \frac{m_c^{\frac{3}{2}}}{3 \gamma^{\frac{1}{2}}} R^{-\frac{1}{2}} + \frac{m_c^{\frac{5}{2}}}{20 \gamma^{\frac{3}{2}}} R^{-\frac{3}{2}} + \order(R^{-\frac{5}{2}}), \\
%\label{eq:J_c_Expansion}
 \bar L_c & = \frac{\pi \gamma}{4} R^2 - \frac{m_c^{\frac{3}{2}}}{3 \gamma^{\frac{1}{2}}} R^{\frac{1}{2}} + \frac{3 m_c^{\frac{5}{2}}}{20 \gamma^{\frac{3}{2}}} R^{-\frac{1}{2}} + \order(R^{-\frac{3}{2}}). \nn
\end{align}
We thus obtain the modified Regge trajectory
\begin{equation}
\label{eq:ReggeExpansion}
 \bar E^2 = 2 \pi \gamma \bar L + \frac{2^\frac{1}{4} 4 \pi^\frac{3}{4}}{3} \gamma^{\frac{1}{4}} \left(m_+^{\frac{3}{2}} + m_-^{\frac{3}{2}} \right) \bar L^{\frac{1}{4}} - \frac{2^\frac{3}{4} \pi^\frac{5}{4}}{10} \gamma^{-\frac{1}{4}} \left(m_+^{\frac{5}{2}} + m_-^{\frac{5}{2}} \right) \bar L^{-\frac{1}{4}} + \order(R^{-1}).
\end{equation}
This gives the next-to-next-to-leading order correction to the Regge trajectory for non-vanishing quark masses. Analogously to \eqref{eq:MasslessReggeTrajectoryShifted}, we define the Regge intercept $a$ as the $\order(L^0)$ shift of this relation, i.e.,
\beq
\label{eq:MassiveReggeTrajectoryShifted}
 E^2 = 2 \pi \gamma (L - a) + C L^{\frac{1}{4}} + \order(L^{- \frac{1}{4}}),
\eeq
with some constant $C$.

For later convenience it is helpful to note that the inclusion of an Einstein-Hilbert term
\beq
\label{eq:S_EH}
 \mathcal{S}_{EH} = - \frac{\alpha}{2} \int_\Sigma \mathcal{R} \sqrt{\betrag{g}}
\eeq
into the action \eqref{eq:Action}, which by the Gau{\ss}-Bonnet theorem is equivalent to the addition of geodesic curvature boundary terms, modifies the subleading term in \eqref{eq:ReggeExpansion}, i.e., \cite{HadaszRog96}
\begin{equation}
\label{eq:EH_Classical}
 \bar E^2 = 2 \pi \gamma \bar L - \frac{4 \pi^\frac{3}{4}}{2^\frac{1}{4}  3} \gamma^{\frac{1}{4}} \left[ \sum_{c \in \pm} \left( \sqrt{m_c^2 + 4 \alpha \gamma} - 2 m_c \right) \sqrt{m_c + \sqrt{m_c^2 + 4 \alpha \gamma}} \right] \bar L^{\frac{1}{4}} + \order(L^{-\frac{1}{4}}).
\end{equation}
It is remarkable that the leading order effects of an Einstein-Hilbert (or geodesic curvature) term and masses at the endpoints occur at the same order. In particular, the coefficient of the sub-leading term has no definite sign.
Furthermore, even for coinciding masses, $m_+ = m_- = m$, the coefficient of the $\order(L^{\frac{1}{4}})$ term does not determine the coefficient of the $\order(L^{-\frac{1}{4}})$ term, unless either $\alpha$ or $m$ are known. But, as we will argue below, $\alpha$ is subject to renormalization ambiguities.

\section{Fluctuations of the rotating string}
\label{sec:Quantization}

Our goal is now to perform a (canonical) quantization of the fluctuations $\vp$ around the classical background $\bar X$, \cf \eqref{eq:X}, i.e., we consider $X = \bar X + \gamma^{-\frac{1}{2}} \varphi$. At second order in $\vp$, i.e., at $\order(\gamma^0)$, the fluctuations parallel to the world-sheet drop out of the bulk part of the action \cite{BRZ}, and analogously, the fluctuations parallel to the boundary drop out of the boundary action. This is analogous to $\vp + \vartheta$ dropping out of the free Lagrangian \eqref{eq:HydrogenH0_0} in our semi-classical treatment of the hydrogen atom. Hence, it is natural to parameterize the fluctuations as
\begin{align}
\label{eq:phi_Parametrization}
 \varphi^a = f_s \begin{pmatrix} 0 \\ 0 \\ 0 \\ 1 \end{pmatrix} + f_p \begin{pmatrix} \tan \sigma \\ - \sin \tau / \cos \sigma \\ \cos \tau / \cos \sigma \\ 0 \end{pmatrix} + f_r \begin{pmatrix} 0 \\ \cos \tau \\ \sin \tau \\ 0 \end{pmatrix}.
\end{align}
Here the \emph{scalar} component $f_s$ describes the fluctuations in the direction perpendicular to the plane of rotation, and the \emph{planar} component $f_p$ describes the fluctuations in the plane of rotation (at least approximately for small $\sigma$). These components are orthonormal to each other and the bulk world-sheet. The \emph{radial} component $f_r$ is orthonormal to the others and the boundary of the world-sheet. This component is only relevant at the boundary, as is obvious from the action \cite{RotatingString}
\begin{align}
\label{eq:S0}
 \sS_0 & = \frac{1}{2} \int_\Sigma \left( {\dot f_p}^2 - {f'_p}^2 - \tfrac{2}{\cos^2 \sigma} f_p^2 + {\dot f_s}^2 - {f'_s}^2 \right) \ud \sigma \ud \tau \nn  \\
 & + \frac{1}{2} \sum_{c \in \pm} \frac{1}{\tan S_c} \int_{\del_c \Sigma} \left( {\dot f_p}^2 + {\dot f_r}^2 + {\dot f_s}^2 + \tfrac{1}{\cos^2 S_c} f_p^2  \right. \nn \\
 & \qquad \qquad \qquad \qquad \left. + (1+2 \tan^2 S_c) f_r^2 + \tfrac{2}{\cos S_c} ( \dot f_p f_r - f_p \dot f_r )  \right) \ud \tau.
\end{align}

Of course, going to higher dimensional target space-time simply amounts to multiplying the number of scalar fields. Furthermore, it should be noted that the string world-sheet is actually curved, \cf \eqref{eq:metric}. This does not matter for the canonical quantization procedure described in this section, but will be important in the discussion of renormalization in the following one.

From the action \eqref{eq:S0}, one obtains the bulk equations of motion (where derivates \wrt $\tau$ are denoted by dots and those \wrt $\sigma$ by primes)
\begin{align}
\label{eq:eom_s}
 - \ddot f_s + f''_s & = 0, \\
\label{eq:eom_p}
 - \ddot f_p + f_p'' - \tfrac{2}{\cos^2 \sigma} f_p & = 0,
\end{align}
supplemented by the boundary conditions
\begin{align}
\label{eq:bc_s}
 - \ddot f_{s}(\pm S_\pm) & = \pm \tan S_\pm f_{s}'(\pm S_\pm), \\
\label{eq:bc_p}
 - \ddot f_{p}(\pm S_\pm) + \tfrac{1}{\cos^2 S_\pm} f_{p}(\pm S_\pm) - \tfrac{2}{\cos S_\pm}  \dot f_{r}(\pm S_\pm) & = \pm \tan S_\pm f'_{p}(\pm S_\pm), \\
\label{eq:bc_r}
 - \ddot f_{r}(\pm S_\pm) + \left( 1 + 2 \tan^2 S_\pm \right) f_{r}(\pm S_\pm) + \tfrac{2}{\cos S_\pm} \dot f_{p}(\pm S_\pm) & = 0.
\end{align}
In fact these boundary conditions can also be interpreted as equations of motion on the boundary, with boundary values of normal derivatives of the bulk fields as sources (on the \rhs of the equations). This point of view was taken in \cite{Wentzell}, where it was shown that the \emph{scalar sector}, i.e., \eqref{eq:eom_s} and \eqref{eq:bc_s}, has a well-posed initial value formulation and causal propagation. It is obvious that the scalar fluctuations decouple, whereas the planar and radial one are coupled. We thus introduce the notation
\[
 f_q = (f_p, f_r)
\]
for the perturbations in the \emph{planar sector}.

In the massless limit, the boundary conditions \eqref{eq:bc_s} for the scalar polarization turn into Neumann boundary conditions, as for the massless string. However, the planar boundary conditions do not converge to the boundary conditions 
of the massless string, \cf Section~\ref{sec:Massless}.

We will canonically quantize this system. A basic ingredient in this is the symplectic form, which is non-standard due to the presence of single time derivative terms in the action \eqref{eq:S0}:
\begin{multline}
\label{eq:SymplecticForm}
 \sigma((f^1, \dot f^1),(f^2, \dot f^2)) = \int_{-S_-}^{S_+} \left( f^1_s \dot f^2_s -\dot f^1_s f^2_s + f^1_p \dot f^2_p - \dot f^1_p f^2_p \right) \\
  + \sum_{c \in \pm} \tfrac{1}{\tan S_c} \left( f^1_s \dot f^2_s -\dot f^1_s f^2_s + f^1_p \dot f^2_p - \dot f^1_p f^2_p + f^1_r \dot f^2_r - \dot f^1_r f^2_r - \tfrac{2}{\cos S_c} (f^1_r f^2_p - f^1_p f^2_r) \right).
\end{multline}
Canonical quantization in such a situation is systematically developed in the appendix to \cite{Fulling}. All our results, in particular on the behavior of symplectically non-normalizable modes, is consistent with the general results derived there.

The basis of the canonical quantization of the system are mode solutions, i.e., solutions of the form
\begin{align*}
 f_{s,n}(\tau, \sigma) & = f_{s,n}(\sigma) e^{- i \omega^s_n \tau}, \\
 f_{q,n}(\tau, \sigma) & = f_{q,n}(\sigma) e^{- i \omega^q_n \tau}.
\end{align*}
The corresponding modes for the bulk equations of motions are
\begin{align}
\label{eq:ModesGeneral_s}
 f_{s, n} & = A \cos \omega^s_n \sigma + B \sin \omega^s_n \sigma, \\
\label{eq:ModesGeneral_p}
 f_{p, n} & = A \left( \omega^q_n \cos \omega^q_n \sigma + \tan \sigma \sin \omega^q_n \sigma \right)  + B \left( \omega^q_n \sin \omega^q_n \sigma - \tan \sigma \cos \omega^q_n \sigma \right).
\end{align}
Setting $B$ ($A$) to zero yields (anti-) symmetric modes, which are realized for coinciding masses $m_+ = m_-$, by symmetry.

One easily checks that both scalar and planar modes always have the lowest non-negative eigenvalues $\omega_0 = 0$, $\omega_1 = 1$, where
\begin{align}
\label{eq:ZeroModes}
 f_{s, 0} & = 1, &
 f_{q, 0} & = (\tan \sigma, 0),  \\
 f_{s, 1} & = \sin \sigma, &
 f_{q, 1} & = (\tfrac{1}{\cos \sigma}, i). \nn
\end{align}
These have a natural geometric interpretation \cite{RotatingString}: The scalar zero mode corresponds to a translation in the direction orthogonal to the plane of rotation and the planar zero mode to a rotation in that plane. The scalar $\omega=1$ mode corresponds to rotations in a plane spanned by $\vec e_3$ and a vector in the plane of rotation, and the planar $\omega=1$ mode to translations in the plane of rotation.\footnote{The phase of the mode determines the corresponding vector in the plane of rotation.} These modes can thus be interpreted as (pseudo-) Goldstone modes for these broken symmetries.

Note that in the planar sector, for coinciding masses, the modes with odd (even) $n$ are (anti-) symmetric, in contrast to the open string case, \cf Section~\ref{sec:Massless}. This is a manifestation of the fact, discussed above, that the planar boundary conditions of the Chodos-Thorn string do not turn into the boundary conditions of the open string in the massless limit. Nevertheless, in both cases the same intercept $a$ will be found.

There are solutions growing linearly in time, associated to the zero modes \eqref{eq:ZeroModes}. One easily checks that they form canonical pairs with the zero modes, when normalized as
\begin{align*}
 f_{s, Q} & = \frac{1}{\sqrt{ \sum_{c \in \pm} (S_c + \cot S_c)}} 1, \\
 f_{s, P} & = \frac{1}{\sqrt{ \sum_{c \in \pm } (S_c + \cot S_c)}} \tau
\end{align*}
and
\begin{align*}
 f_{q, \theta} & = \frac{1}{\sqrt{\sum_{c \in \pm} (S_c + \frac{\sin 2 S_c}{1+\sin^2 S_c})}} (\tan \sigma, 0), \\
 f_{q, \lambda} & = - \frac{1}{\sqrt{\sum_{c \in \pm} (S_c + \frac{\sin 2 S_c}{1+\sin^2 S_c})}} (\tau \tan \sigma, \mp \tfrac{2 \sin S_\pm}{1+\sin^2 S_\pm} ),
\end{align*}
i.e.,\footnote{Here and in the following, we identify a solution $f$ with its Cauchy data $(f, \dot f)$.}
\[
 \sigma(f_{s, Q}, f_{s, P}) = \sigma(f_{p, \theta}, f_{p, \lambda}) = 1.
\]
We note the unusual sign of the linearly growing mode $f_{q, \lambda}$, which was also found for the analogous $f_\lambda$ mode in our semi-classical treatment of the hydrogen atom in Section~\ref{sec:Hydrogen}.
It is natural to interpret $(f_{s, Q}, f_{s, P})$, or rather the coefficients of these modes, as a pair of position and momentum perpendicular to the plane of rotation and $(f_{q, \theta}, f_{q, \lambda})$ as a pair of angle and angular momentum in the $1-2$ plane. This will be corroborated below.

For the planar sector, there is even a linearly growing solution associated to the $n=1$ mode. To be precise, we define
\begin{align*}
 f_{q, Q} & = \frac{1}{\sqrt{2 \sum_{c \in \pm} (S_c + \cot S_c)}} (\tfrac{1}{\cos \sigma}, i) e^{- i \tau}, \\
 f_{q, P} & = \frac{1}{\sqrt{2 \sum_{c \in \pm} (S_c + \cot S_c)}} (\tfrac{\tau}{\cos \sigma} + i \cos \sigma, i \tau - 1) e^{- i \tau} + u f_{q, Q},
\end{align*}
with
\[
 u = - \frac{i}{4} \frac{\sum_{c \in \pm} (3 S_c + 4 \cot S_c) - \cos(S_+ - S_-) \sin (S_+ + S_-)}{\sum_{c \in \pm} (S_c + \cot S_c)}.
\]
We then have
\begin{align}
\label{eq:n1Normalization}
 \sigma(\overline{f_{q, Q}}, f_{q, P}) & = 1, &
 \sigma(f_{q, Q}, f_{q, P}) =
 \sigma(\overline{f_{q, Q}}, f_{q, Q}) =
 \sigma(\overline{f_{q, P}}, f_{q, P}) & = 0.
\end{align}
Hence, $(\overline{f_{q, Q}}, f_{q, P})$ and $(f_{q, Q}, \overline{f_{q, P}})$ are pairs of canonically conjugate variables. The linearly growing modes $f_{q, P}$, $\overline{f_{q,P}}$ correspond to a uniform movement in the plane of rotation. This suggest that we should view these modes as positions and momenta in the plane of rotation, \cf also below.

The scalar modes with $n \geq 1$ and the planar modes with $n \geq 2$ are normalized symplectically as
\beq
\label{eq:Normalization}
 \sigma(\overline{f_{r, n}}, f_{r', n'}) = - i \delta_{r r'} \delta_{n n'},
\eeq
where $r \in \{ s, q \}$.
The normalization \eqref{eq:Normalization} amounts to
\begin{align}
\label{eq:Normalization_s}
 \delta_{n m} & = \left( \omega^s_n + \omega^s_m \right) \left[ \int_{-S_-}^{S_+} f_{s,n} f_{s,m} + \sum_{c \in \pm} \tfrac{1}{\tan S_c} f_{s, n} f_{s, m} \right], \\
 \delta_{n m} & = \left( \omega^q_n + \omega^q_m \right) \left[ \int_{-S_-}^{S_+} f_{p,n} f_{p,m} + \sum_{c \in \pm} \tfrac{1}{\tan S_c} \left( f_{p, n} f_{p, m} - f_{r, n} f_{r, m} \right)  \right] \nonumber \\
\label{eq:Normalization_p}
 & + \sum_{c \in \pm} \tfrac{2 i}{\sin S_c} \left( f_{r,n} f_{p,m} + f_{p,n} f_{r,m} \right).
\end{align}
for $n, m > 0$.

In order to prepare for the canonical quantization, we write
\begin{align}
\label{eq:phi_Expansion_s}
 \phi_s & = \sum_{n \in \N_s} \left( a_{s, n} f_{s, n} + \text{h.c.} \right) + Q_s f_{s,Q} + P_s f_{s,P} \\
\label{eq:phi_Expansion_q}
 \phi_q & = \sum_{n \in \N_q} \left( a_{q, n} f_{q, n} + \text{h.c.} \right) + \theta f_{q, \theta} + \lambda f_{q, \lambda} + \left( Q_{q} f_{q, Q} + P_q f_{q, P} + \text{h.c.} \right),
\end{align}
where
\begin{align*}
 \N_s & = \{ n \geq 1 \}, &
 \N_q & = \{ n \geq 2 \},
\end{align*}
and the coefficients $Q_s, P_s, \theta, \lambda$ are real. One then finds, for the expansion of the energy, \cf \eqref{eq:Momentum},
\begin{align}
 E & = \bar E + \sqrt{\gamma} \left[ \int_{-S_-}^{S_+} \tan \sigma \dot f_p(\sigma) \ud \sigma + \sum_{c \in \pm} \left( \dot f_p(c S_c) + \tfrac{2}{\cos S_c} f_r(c S_c) \right) \right] + \order(\gamma^0), \nn \\
 & = \bar E + \sqrt{\gamma} \sqrt{\sum\nolimits_{c \in \pm} (S_c + \tfrac{\sin 2 S_c}{1+\sin^2 S_c})} \sigma(f_{q, \theta},  \phi) + \order(\gamma^0) \nn \\
\label{eq:E_1st_order}
 & = \bar E + \sqrt{\gamma} \sqrt{\sum\nolimits_{c \in \pm} (S_c + \tfrac{\sin 2 S_c}{1+\sin^2 S_c})} \lambda + \order(\gamma^0).
\end{align}
Similarly, one obtains for the angular momentum and the momenta, \cf \eqref{eq:Momentum}, \eqref{eq:AngularMomentum},
\begin{align}
\label{eq:L_1st_order}
 L_{1, 2} & = \bar L_{1, 2} + \sqrt{\gamma} R \sqrt{\sum\nolimits_{c \in \pm} (S_c + \tfrac{\sin 2 S_c}{1+\sin^2 S_c})} \lambda + \order(\gamma^0), \\
\label{eq:P3_1st_order}
 P^3 & = \sqrt{ \gamma  \sum\nolimits_{c\in \pm} (S_c + \cot S_c)} P_s + \order(\gamma^0) = \sqrt{\bar E/R} P_s + \order(\gamma^0), \\
\label{eq:P1_1st_order}
 P^1 & = \sqrt{2 \gamma \sum\nolimits_{c\in \pm} (S_c + \cot S_c)} \Im \bar P_q + \order(\gamma^0) = \sqrt{2 \bar E/R} \Im \bar P_q + \order(\gamma^0), \\
\label{eq:P2_1st_order}
 P^2 & = \sqrt{2 \gamma \sum\nolimits_{c\in \pm} (S_c + \cot S_c)} \Re P_q + \order(\gamma^0) = \sqrt{2 \bar E/R} \Re P_q + \order(\gamma^0).
\end{align}
This supports the identification of the modes $f_{q, \lambda}$, $f_{s, P}$, $f_{q, P}$ with (angular) momenta discussed above.

Canonical quantization now proceeds as follows: One introduces annihilation and creation operators $\hat a_{r, n}$, $\hat{a}^{*}_{r, n}$ for $r \in \{ s, q \}$, $n \in \N_r$, fulfilling
\[
 [ \hat a_{r, n}, \hat a^{*}_{r', n'} ] = \delta_{r r'} \delta_{n n'}.
\]
Furthermore, one introduces position operators $\hat Q_s, \hat \theta, \hat Q_q, \hat Q_q^*$ and momenta $\hat P_s, \hat \lambda, \hat P_q, \hat P_q^*$ with commutation relations\footnote{The complex positions $\hat Q_q$ can be represented on $L^2(\R^2)$ as $\hat Q_q = \frac{1}{\sqrt{2}} ( Q_1 + i Q_2 )$, and analogously for the momenta.}
\begin{align*}
 [\hat Q_s, \hat P_s] & = i, &
 [\hat \theta, \hat \lambda] & = i, &
 [\hat Q_q^*, \hat P_q] & = i, &
 [\hat Q_q, \hat P_q] & = 0.
\end{align*}
Then one replaces the coefficients in \eqref{eq:phi_Expansion_s}, \eqref{eq:phi_Expansion_q} by the hatted corresponding operators. 
The fulfillment of the canonical equal time commutation relations then follows from completeness of the modes. Mathematically, this is expressed by the fact that the Cauchy data of
\[
 \{ f_{s, n}, \overline{f_{s, n}} \}_{n \in \N_s} \cup \{ f_{s, Q}, f_{s, P} \} \cup \{ f_{q, n}, \overline{f_{q,n}} \}_{n \in \N_q} \cup \{ f_{q,Q}, f_{q,P}, \overline{f_{q,Q}}, \overline{f_{q,P}}, f_{q,\theta}, f_{q,\lambda} \}
\]
are a basis of a Krein space with indefinite inner product given by
\[
 [f|g] = i \sigma(\bar{f}, g),
\]
or, more precisely, that
\begin{multline*}
 \sum_{r \in \{s, q \}} \sum_{n \in \N_r} \left( |f_{r, n}][f_{r, n}| - |\overline{f_{r, n}}][\overline{f_{r, n}}| \right) + i |f_{s, Q}][f_{s, P}| - i |f_{s, P}][f_{s, Q}| + i |f_{q, \theta}][f_{q, \lambda}| - i |f_{q, \lambda}][f_{q, \theta}| \\
  + i |f_{q, Q}][f_{q, P}| - i |f_{q, P}][f_{q, Q}| + i |\overline{f_{q, Q}}][\overline{f_{q, P}}| - i |\overline{f_{q, P}}][\overline{f_{q, Q}}|  = \1.
\end{multline*}
This is due to the fact that the Hamiltonian on this Krein space is Krein self-adjoint, definitizable, and regular at infinity \cite{Langer82} and has a real spectrum. As proofs of these statements lie outside of the main interest of this paper, we omit them.

Omitting the positions and momenta (this will be justified below) we thus have quantum fields $\phi_s$, $\phi_q$ with two-point functions
\begin{align}
\label{eq:2pt_s}
 w_s(x; x') & \defeq \bra{\Omega} \phi_s(x) \phi_s(x') \ket{\Omega} = \sum_{n \in \N_s} f_{s, n}(x) \overline{f_{s, n}}(x'), \\
%\label{eq:2pt_p}
 w_q(x; x') & \defeq \bra{\Omega} \phi_q(x) \phi_q(x') \ket{\Omega} = \sum_{n \in \N_q} f_{q, n}(x) \overline{f_{q, n}}(x'), \nn
\end{align}
where for the planar sector one has to take into account also the radial component at the boundary.

\section{The world-sheet Hamiltonian and the target space energy}
\label{sec:Energy}

The free world-sheet Hamiltonian for the fluctuations around the rotating string solutions has been derived in \cite{RotatingString}:
\begin{align}
 H^0 & = \frac{1}{2} \int_{-S_-}^{S_+} \left( \dot \phi_p^2 + {\phi'_p}^2 + \tfrac{2}{\cos^2 \sigma} \phi_p^2 + \dot \phi_s^2 + {\phi'_s}^2 \right) \ud \sigma \nn \\
\label{eq:H_0}
 & + \frac{1}{2} \sum_{c \in \pm} \frac{1}{\tan S_c} \left( \dot \phi_p^2 + \dot \phi_r^2 - \tfrac{1}{\cos^2 S_c} \phi_p^2 - (1+2 \tan^2 S_c) \phi_r^2 + \dot \phi_s^2  \right).
\end{align}
With \eqref{eq:phi_Expansion_s}, \eqref{eq:phi_Expansion_q}, this can formally be written as
\begin{equation}
\label{eq:H_0_Expansion}
 H^0 = \frac{1}{2} \sum_{r \in \{ s, q \}} \sum_{n \in \N_r} n \left( \hat a_{r, n} \hat a_{r,n}^* + \hat a_{r, n}^* \hat a_{r,n} \right) - \frac{1}{2} \hat \lambda^2 + \frac{1}{2} \hat P_s^2 + i \hat P_q {\hat Q_q}^* - i {\hat P_q}^* \hat Q_q + {\hat P_q}^* \hat P_q.
\end{equation}
We note the similarity of this expression, regarding the presence of the negative energy $\lambda^2$ mode, with the free Hamiltonian \eqref{eq:H0_Hydrogen} derived for the semi-classical hydrogen atom.

To understand the significance of this free world-sheet Hamiltonian, we note the relation
\begin{equation}
\label{eq:Relation_H_E}
 H = R E^q - L_{1,2}^q,
\end{equation}
analogous to \eqref{eq:Relation_H_E_Hydrogen} in the semi-classical hydrogen atom,
between the full world-sheet Hamiltonian $H$ and the quantum corrections $E^q$ and $L^q_{1,2}$ to the (target space) energy and angular momentum. The latter are defined by the split
\begin{align*}
 E & = \bar E + E^q, &
 L_{1,2} & = \bar L_{1,2} + L_{1,2}^q,
\end{align*} 
into the classical and the $\vp$ dependent parts.
The relation \eqref{eq:Relation_H_E} is a consequence of the fact that $H$ generates translations in the world-sheet time $\tau$, whereas $E^q$ generates translations in the target space time $X^0$. The factor $R$ is due to the relation between the two, \cf \eqref{eq:X}. Furthermore, the time evolution generated by $H$ acts on the coefficient of the basis vectors $v_p, v_r, v_s$, \cf \eqref{eq:phi_Parametrization}. The first two of these rotate, which is seen by the time evolution generated by $E^q$. To correct this, the generator of rotations has to be added. The relation \eqref{eq:Relation_H_E} has been already checked to first order in the perturbation, \cf \eqref{eq:E_1st_order} and \eqref{eq:L_1st_order}, as the Hamiltonian $H$ does not have a first order term. It can also easily be checked that the second order term on the right hand side coincides with the free Hamiltonian \eqref{eq:H_0}.

Furthermore, the classical solution breaks the time translation invariance to discrete translations $X^0 \mapsto X^0 + 2 \pi R$. These correspond to world-sheet translations $\tau \mapsto \tau + 2 \pi$. Hence,
\[
 E^q = \tfrac{1}{R} H \mod \tfrac{1}{R}.
\]
With \eqref{eq:Relation_H_E}, it follows that $L^q_{1,2}$ must have spectrum in the integers, as expected for an angular momentum operator. By \eqref{eq:L_1st_order}, this implies that $\hat \lambda$ has a discrete spectrum with eigenvalue $0$. In the following, we are only considering such eigenstates. In particular, this means that the first order corrections to $L_{1,2}$ and $E$ vanish.

Let us thus consider the second order correction to $E^2$. Using that the first order variation $\delta^1 E$ of $E$ vanishes (we write $E = \bar E + \sum_k \delta^k E$, with $k$ denoting the order of the perturbation $\vp$ appearing in $\delta^k E$), we have, by \eqref{eq:E_c_Expansion} and \eqref{eq:Relation_H_E},
\[
 \delta^2 E^2 = 2 \bar E \delta^2 E = 2 \pi \gamma ( \delta^2 L_{1,2} + H^0 ) + \order(R^{-\frac{3}{2}}),
\]
where we also used that, by \eqref{eq:metric}, \eqref{eq:S_Condition} and \eqref{eq:Momentum}, \eqref{eq:AngularMomentum}, $R \delta^2 E$ and $\delta^2 L_{1,2}$ are classically of $\order(R^0)$. Plugging this into \eqref{eq:ReggeExpansion}, we find, with $E^2 = \bar E^2 + \delta^2 E^2$ and $L = \bar L + \delta^2 L$,
\beq
\label{eq:E2Intercept}
 E^2 = 2 \pi \gamma ( L + H^0 ) + \frac{4 \pi}{3} \gamma^{\frac{1}{2}} \left(m_+^{\frac{3}{2}} + m_-^{\frac{3}{2}} \right) \left( \frac{2}{\gamma \pi} \right)^{\frac{1}{4}} L^{\frac{1}{4}} + \order(L^{-\frac{1}{4}}).
\eeq
Comparison with \eqref{eq:MassiveReggeTrajectoryShifted} shows that we can determine the intercept $a$ by computing the $\order(R^0)$ contribution of the vacuum expectation value of the free Hamiltonian $H^0$.\footnote{As we will see below, the expectation value of $H^0$ has a term of $\order(R^{\frac{1}{2}})$, due to logarithmic divergences. This, however, is a renormalization ambiguity, corresponding to a geodesic curvature boundary term affecting the coefficient of the $\order(L^{\frac{1}{4}})$ term, \cf \eqref{eq:EH_Classical}.}

Let us discuss the influence of the $P^2$ terms in \eqref{eq:H_0_Expansion}. Using \eqref{eq:P3_1st_order}, \eqref{eq:P1_1st_order}, \eqref{eq:P2_1st_order}, and \eqref{eq:E_c_Expansion}, we see that the leading order contribution to $E^2$ from these terms is
\[
 E^2 = \frac{\pi \gamma R}{\bar E} P_i^2 = P_i^2 + \order(R^{- \frac{3}{2}}),
\]
as one would expect. For the determination of the intercept, this spatial momentum contribution to the energy should of course be neglected. Furthermore, one can easily see that the $PQ$ terms in \eqref{eq:H_0_Expansion} are the center of mass contribution to the angular momentum $- L_{1,2}$. By \eqref{eq:Relation_H_E}, such a term has to be expected in $H$, as, for a non-zero spatial momentum, one can, by a translation, change the angular momentum $L_{1,2}$ without changing the energy. \eqref{eq:Relation_H_E} can thus only be correct if this is compensated in $H$. For the determination of the Regge trajectory, one has of course to consider a vanishing center of mass contribution to the angular momentum. Hence, all but the first term on the \rhs of \eqref{eq:H_0_Expansion} should be neglected for the determination of the intercept.

\section{Renormalizing the world-sheet Hamiltonian: The massless string}
\label{sec:Massless}

As for the massive string the evaluation of the renormalized world-sheet Hamiltonian has to be performed numerically, we begin by discussing the massless case first, where an analytic treatment is possible. This has the advantage that the tools necessary for local renormalization can be introduced in a more transparent context.

In the context of the massless string, it is advantageous\footnote{The advantage is that one can write the symmetric and anti-symmetric eigenmodes in a uniform notation.} to choose coordinates $\sigma \in (0, \pi)$ such that the metric, the scalar curvature, and the equation of motion for the planar polarization are given by
\begin{align}
\label{eq:MetricMassless}
 g_{\mu \nu} & = R^2 \sin^2 \sigma \eta_{\mu \nu}, \\
\label{eq:RMassless}
 \cR & = \frac{2}{R^2 \sin^4 \sigma}, \\
 - \ddot f_p & = - f''_p + \tfrac{2}{\sin^2 \sigma} f_p.
\end{align}
The trajectory $\tau \mapsto (\tau, s)$, with $s$ fixed, has the geodesic curvature
\beq
\label{eq:GeodesicCurvatureMassless}
 \kappa_s = - \frac{\cot s}{R \sin s}.
\eeq
From the massless boundary condition
\beq
\label{eq:bc}
 \sqrt{\betrag{g}} g^{1 \mu} \del_\mu X = 0,
\eeq
one derives the boundary conditions
\begin{align}
\label{eq:bc_s_massless}
 0 & = f_{s}'(0) = f_{s}'(\pi), \\
\label{eq:bc_p_massless}
 0 & = f_p(0) = f_p(\pi) = f_p'(0) = f_p'(\pi)
\end{align}
for the scalar and the planar polarization, as shown in Appendix~\ref{app:bc}.

The operators $\Delta_s = - \del_\sigma^2$, $\Delta_p = - \del_\sigma^2 + \frac{2}{\sin^2 \sigma}$ on $L^2([0,\pi])$, on the domain $C^2([0, \pi])$ with boundary conditions \eqref{eq:bc_s_massless}, \eqref{eq:bc_p_massless}, are essentially self-adjoint, so they admit a unique self-adjoint extension.\footnote{For $\Delta_s$ this is clear. $\Delta_p$ is obviously symmetric. It thus remains to show that the deficiency indices vanish. The generic solution to $\Delta_p f = \pm i f$ is
\[
 f(x) = C_1 \sqrt{\sin \sigma} P_{(i \pm 1)/\sqrt{2} - 1/2}^{3/2}(\cos \sigma) + C_2 \sqrt{\sin \sigma} Q_{(i \pm 1)/\sqrt{2} - 1/2}^{3/2}(\cos \sigma).
\]
It is easy to see that there are no normalizable solutions of this form.} Defining $\N_N = \{ n \in \N | n \geq N \}$, these have spectrum $\N_0$, $\N_2$, with normalized (\wrt the $L^2$ inner product, not the symplectic form \eqref{eq:SymplecticForm}) eigenvectors
\begin{align*}
 f_{s, n} & = \tfrac{\sqrt{2}}{\sqrt{\pi}} \cos n \sigma, \\
 f_{p, n} & = \tfrac{\sqrt{2}}{\sqrt{\pi (n^2-1)}} \left( n \cos n \sigma - \cot \sigma \sin n \sigma \right).
\end{align*}
In the massless case, the planar $n=0$ and $n=1$ mode are thus absent, as already noted in \cite{BakerSteinke01, ZayasSonnenscheinVaman}.
The scalar zero mode corresponds to translations perpendicular to the plane of rotation. There is also an associated momentum. For the purposes of the calculation of the Regge intercept, we want to fix the spatial momentum, so we do not consider the zero modes in the following. The usual canonical quantization then yields quantum fields $\phi_s$, $\phi_p$ with two-point functions
\begin{align}
\label{eq:2pt_s_massless}
 w_s(x; x') & \defeq \bra{\Omega} \phi_s(x) \phi_s(x') \ket{\Omega} = \sum_{n \geq 1} \tfrac{1}{2 n} f_{s, n}(\sigma) f_{s, n}(\sigma') e^{- i n (\tau - \tau' - i \eps)}, \\
\label{eq:2pt_p_massless}
 w_p(x; x') & \defeq \bra{\Omega} \phi_p(x) \phi_p(x') \ket{\Omega} = \sum_{n \geq 2} \tfrac{1}{2 n} f_{p, n}(\sigma) f_{p, n}(\sigma') e^{- i n (\tau - \tau' - i \eps)}.
\end{align}

The canonical quantization scheme in particular implies that the physical fluctuations are represented on a positive definite Fock space.

The free Hamiltonian corresponding to the free action~\eqref{eq:S0} is
\begin{equation*}
%\label{eq:H_0}
 H^0 = \frac{1}{2} \int_{0}^{\pi} \left( \dot \phi_p^2 + {\phi'_p}^2 + \tfrac{2}{\sin^2 \sigma} \phi_p^2 + \dot \phi_s^2 + {\phi'_s}^2 \right) \ud \sigma.
\end{equation*}
We see that the planar and the scalar polarization decouple. Let us first concentrate on the scalar sector. Formally, the vacuum expectation value is given by
\[
 \langle H^0_s \rangle = \frac{1}{2} \sum_{n \geq 1} n.
\]
This sum is of course quadratically divergent. As long as one does not impose some conditions on the renormalization prescription, one can obtain any result.
The renormalization prescription that we are going to employ is based on the framework of locally covariant field theory \cite{HollandsWaldWick}, where the renormalization is performed locally, by using the local geometric data. In that framework, the expectation value of Wick squares (possibly with derivatives) is determined as follows:
\[
 \bra{\Omega} (\nabla^\alpha \phi \nabla^\beta \phi)(x) \ket{\Omega} = \lim_{x' \to x} \nabla^\alpha {\nabla'}^\beta \left( w(x; x') - h(x; x') \right)
\]
Here $\alpha, \beta$ are multiindices, $w$ is the two-point function in the state $\Omega$, defined as on the \lhs of \eqref{eq:2pt_s_massless}, \eqref{eq:2pt_p_massless}, and $h$ is a distribution which is covariantly constructed out of the local geometric data, the \emph{Hadamard parametrix}. Importantly, for physically reasonable states (ground states in particular), the difference $w - h$ is smooth, so that the above coinciding point limit exists and is independent of the direction from which $x'$ approaches $x$. This method has been reliably used for the computation of Casimir energies and vacuum polarization, \cf \cite{HollandsWaldReview, DappiaggiNosariPinamonti14, Current} for example.

For our purposes, it is advantageous to perform the limit of coinciding points from the time direction, i.e., we take $x = (\tau, \sigma)$, $x' = (\tau + t, \sigma)$, and $t \to +0$. Performing the summation in \eqref{eq:2pt_s}, we find 
\[
 \frac{1}{2} ( \del_0 \del_0' + \del_1 \del_1') w_s(x; x') = - \frac{1}{2 \pi (t+i \eps)^2} - \frac{1}{24 \pi} + \order(t).
\] 
For a minimally coupled scalar field with a variable mass $m^2(x)$ in two dimensional space-time, the Hadamard parametrix is given by (see, e.g., \cite{DecaniniFolacci08})
\[
 h(x; x') = - \frac{1}{4 \pi} \left( 1 + \frac{1}{2} m^2(x) \rho(x, x') + \order((x-x')^3) \right) \log \frac{\rho_\eps(x,x')}{\Lambda^2},
\]
where $\rho$ is the \emph{Synge world function}, i.e., $\frac{1}{2}$ times the squared (signed) geodesic distance of $x$ and $x'$, \cf \cite{Poisson03}, and $\Lambda$ is a length scale (the ``renormalization scale''). For the local covariance, it is crucial that $\Lambda$ is fixed and does not depend on any geometric data \cite{HollandsWaldWick}. Inside of the logarithm, the world function is equipped with an $i \eps$ prescription as follows:
\[
 \rho_\eps(x, x') = \rho(x, x') + i \eps (\tau - \tau').
\]

For the scalar part, the mass term is absent. The world function can be Taylor expanded in coordinates around coinciding points as \cite{OttewillWardell}
\begin{align*}
 \rho(x,x') & = \tfrac{1}{2} g_{\mu \nu}(x) \Delta x^\mu \Delta x^\nu + A_{\mu \nu \lambda}(x) \Delta x^\mu \Delta x^\nu \Delta x^\lambda + B_{\mu \nu \lambda \rho}(x) \Delta x^\mu \Delta x^\nu \Delta x^\lambda \Delta x^\rho, \\
 A_{\mu \nu \lambda} & = - \tfrac{1}{4} \del_{(\mu} g_{\nu \lambda)}, \\
 B_{\mu \nu \lambda \rho} & = \tfrac{1}{12} \del_{(\mu} \del_{\nu} g_{\lambda \rho)} - \tfrac{1}{24} g^{\sigma \tau} \left( \tfrac{1}{4} \del_\sigma g_{(\mu \nu} \del_{|\tau|} g_{\lambda \rho)} - \del_{\sigma} g_{(\mu \nu} \del_{\lambda} g_{\rho) \tau} + \del_{(\mu} g_{\nu | \sigma|} \del_{\lambda} g_{\rho) \tau} \right),
\end{align*}
where $\Delta x = x - x'$. One thus finds, for a metric of the form $g_{\mu \nu} = f(\sigma) \eta_{\mu \nu}$,
\begin{multline*}
 \rho = \tfrac{1}{2} f(\sigma) \left( - \Delta \tau^2 + \Delta \sigma^2 \right) + \tfrac{1}{4} f'(\sigma) \Delta \tau^2 \Delta \sigma - \tfrac{1}{96} f(\sigma)^{-1} f'(\sigma)^2 \Delta \tau^4 \\ + \left( \tfrac{1}{48} f(\sigma)^{-1} f'(\sigma)^2 - \tfrac{1}{12} f''(\sigma) \right) \Delta \tau^2 \Delta \sigma^2 + \order(\Delta x^5, \Delta \sigma^3),
\end{multline*}
and hence,
for the coinciding point limit from the time direction,\footnote{Here and in the following, $\order(t)$ also includes terms of the form $t \log t$.} 
\[
 \frac{1}{2} ( \del_0 \del_0' + \del_1 \del_1') h_s = - \frac{1}{2 \pi (t+i \eps)^2} + \frac{1}{32 \pi} \frac{f'^2}{f^2} - \frac{1}{48 \pi} \frac{f''}{f} + \order(t).
\]
In the special case of the metric \eqref{eq:MetricMassless}, this yields
\beq
\label{eq:ParametrixEnergyScalarMassless}
 \frac{1}{2} ( \del_0 \del_0' + \del_1 \del_1') h_s = - \frac{1}{2 \pi (t+i \eps)^2} + \frac{1}{12 \pi \sin^2 \sigma} - \frac{1}{24 \pi} + \order(t).
\eeq
For the scalar contribution to the energy density, we thus obtain
\beq
\label{eq:H0_s}
 \VEV{H^0_s(\sigma)} = - \frac{1}{12 \pi \sin^2 \sigma}.
\eeq
This is locally finite, but diverges in a non-integrable fashion at the boundaries. As discussed below, this term may be absorbed in the renormalization freedom of the planar contribution. Alternatively, one recognizes it as a multiple of $\cR \sqrt{-g}$, \cf \eqref{eq:MetricMassless}, \eqref{eq:RMassless}, which can be absorbed in a geodesic curvature boundary counterterm, \cf also the discussion below.

Also the two-point function of the planar part can be computed explicitly. Evaluating the sums in \eqref{eq:2pt_p_massless}, one obtains, \cf Appendix~\ref{app:2pt},
\begin{multline}
\label{eq:2pt_p_eval}
 \tfrac{1}{2} ( \del_0 \del_0' + \del_1 \del_1' + \tfrac{2}{\sin^2 \sigma}) w_p(x; x') \\ = - \frac{1}{2 \pi} \left[ \frac{1}{(t+i \eps)^2} + \frac{1}{2 \sin^2 \sigma} \log \frac{-(t+i\eps)^2}{4 \sin^2 \sigma} + \frac{3}{2 \sin^2 \sigma} + \frac{1}{12} \right] + \order(t).
\end{multline}
For the parametrix, we note that given the metric \eqref{eq:MetricMassless}, the mass square which is implicit in the free action
is
\[
 m^2 = \frac{2}{R^2 \sin^4 \sigma},
\] 
so that we obtain
\begin{multline}
\label{eq:h_p}
 \tfrac{1}{2} \left( \del_0 \del_0' + \del_1 \del_1' + \tfrac{2}{\sin^2 \sigma} \right) h_p \\
 = - \frac{1}{2 \pi} \left[ \frac{1}{(t+i \eps)^2} + \frac{1}{2 \sin^2 \sigma} \log \frac{- (t+i \eps)^2 R^2 \sin^2 \sigma}{\Lambda^2} + \frac{1}{3 \sin^2 \sigma} + \frac{1}{12} \right] + \order(t).
\end{multline}
Hence, for the planar contribution to the energy density, we find
\[
 \VEV{H^0_p(\sigma)} = - \frac{1}{2 \pi \sin^2 \sigma} \log \frac{\Lambda}{2 R \sin^2 \sigma} - \frac{7}{12 \pi \sin^2 \sigma}.
\]
In the last term, we have the same non-integrable divergence that we already found in \eqref{eq:H0_s}. However, we see that both these terms can be absorbed in a change of the scale $\Lambda$. Noting that $\frac{1}{\sin^2 \sigma} = \frac{1}{2} \sqrt{\betrag{g}} \cR$, this corresponds to an Einstein-Hilbert counterterm. In fact, the most general redefinition of a parametrix that affects Wick powers with up to two derivatives is
\[
%\label{eq:h_Redefinition}
  h \mapsto h + c_0 + c_1 \cR \rho + c_2 m^2 \rho. 
\]
This has no effect on the scalar contribution to the energy density and its effect on the planar contribution is exactly corresponding to a finite renormalization of the Einstein-Hilbert term. Our final expression for the local energy density in $D$ dimensional target space is thus
\beq
\label{eq:H0_final}
 \VEV{H^0(\sigma)} = - \frac{1}{2 \pi \sin^2 \sigma} \log \frac{\Lambda}{R \sin^2 \sigma}.
\eeq

The final expression \eqref{eq:H0_final} still contains a non-integrable singularity at the boundaries. We recall that near Dirichlet boundaries, the energy density of a massive scalar field in two space-time dimensions behaves as
\[
 \eps \sim - \frac{m^2}{2 \pi} \log \frac{\lambda}{m d},
\]
with $d$ the distance to the boundary, \cf \cite{GribMamayevMostepanenko} for example. In view of this and the divergence of $m^2$ near the boundary, a divergence as in the second term in \eqref{eq:H0_final} has to be expected. Such non-integrable divergences near boundaries are a well-known phenomenon \cite{DeutschCandelas}, in particular in space-time dimensions larger than two. For the treatment of our singularity, we follow the approach proposed in \cite{DowkerKennedy78}, i.e., to introduce boundary conterterms. Concretely, one performs the integration of the energy density only up to a distance $d$ to the boundary and introduces a $d$-dependent local counterterm on this boundary.
We denote by $s$ the value of $\sigma$ at which this shifted boundary resides.
In the spirit of locally covariant field theory, a boundary counterterm may only depend on the boundary geometric data and the proper distance $d_s = 2 R \sin^2 \tfrac{s}{2}$ to the boundary. More precisely, it should be of the form
\beq
\label{eq:CountertermGeneral}
 \sqrt{\betrag{h^s}} p(d_s^{-1}, \log d_s / \Lambda_{\bd}, \kappa_s, \mathcal{R}(s), m^2(s)),
\eeq
with $h^s$ and $\kappa_s$ the induced metric and geodesic curvature on the boundary, $p$ a polynomial (which may also contain normal derivatives of $\mathcal{R}$ and $m^2$), and $\Lambda_\bd$ a renormalization scale. We compute
\[
 \int_{s}^{\pi-s} \frac{1}{\sin^2 \sigma} \log \frac{\Lambda}{R \sin^2 \sigma} \ud \sigma = - 4 s + 2 \pi + 2 \cot s \log \frac{\Lambda}{e^2 R \sin^2 s}.
\]
The only way to cancel the divergence in the last term with a counterterm of the form \eqref{eq:CountertermGeneral} is to add the counterterm
\[
 \frac{1}{\pi} \sqrt{ \betrag{h^s}} \kappa_s \log \frac{e^2 d_s}{\Lambda},
\]
\cf \eqref{eq:GeodesicCurvatureMassless}.
It is important to note that the scale $\Lambda_\bd$ in the logarithm is fixed by the renormalization scale $\Lambda$, so that there is no renormalization ambiguity (a change in $\Lambda$ would lead to a non-integrable divergence of the energy density, unless compensated by a change of $\Lambda_\bd$).
Similar (geodesic curvature) boundary counterterms for open strings were also used in \cite{BakerSteinke02, HellermanSwanson} for the calculation of the energy.
Hence, for the renormalized total energy, we finally obtain
\[
 \VEV{H^0_{\mathrm{ren}}} = - 1,
\]
which, by \eqref{eq:a_H0}, yields the intercept \eqref{eq:a1}.\footnote{\label{ft:BS}We remark that omitting the factor $\sin^2 \sigma$ in the logarithm in the planar parametrix \eqref{eq:h_p} (such a modification singles out a preferred parametrization of the world-sheet and corresponds to the regularization performed in \cite{BakerSteinke02}) leads to subtracting $\frac{1}{2}$ from the intercept.}

\section{Renormalizing the world-sheet Hamiltonian: The massive string}
\label{sec:Massive}

The first term on the \rhs of \eqref{eq:H_0_Expansion} naturally decomposes into a scalar and a planar contribution,
\[
 H^0 = H^0_s + H^0_q.
\]
Let us first concentrate on the scalar contribution $H^0_s$.
Formally, its vacuum expectation value is given by
\begin{equation}
\label{eq:H_s_formal}
 \langle H^0_s \rangle = \frac{1}{2} \sum_{n \geq 1} \omega^s_n.
\end{equation}
To give meaning to this divergent series, we again perform a local renormalization, as for the massless case treated in the previous section. A major difference to that case is the presence of a boundary term.
Let us start by considering the bulk. According to \eqref{eq:2pt_s}, the scalar two-point function $w_s$ is given by
\[
 w_s(\tau, \sigma; \tau', \sigma') = \sum_{n \geq 1} f_{s, n}(\sigma) f_{s, n}(\sigma') e^{-i \omega^s_n (\tau - \tau')}.
\]
Note that, as discussed above, the contribution of the zero mode is suppressed.
As above, we perform the coinciding point limit from the time direction, i.e., we take $x' = (\tau + t, \sigma)$, where $x = (\tau, \sigma)$ and $t \to +0$. We then obtain
\[
 \frac{1}{2} ( \del_0 \del_0' + \del_1 \del_1') w_s(x; x') = \frac{1}{2} \sum_{n \geq 1} (\omega^s_n c_n^2)^2 e^{i \omega^s_n (t+i \eps)},
\]
where $c^s_n$ are the normalization constants for the scalar modes \eqref{eq:ModesGeneral_s} such that \eqref{eq:Normalization_s} holds.
Using the asymptotic form of $\omega^s$ (for $S_+ = S_-$, this was proven in \cite{Wentzell})
\begin{equation}
\label{eq:omega_s_asymptotic}
 \omega^s_n = \frac{(n-1) \pi}{S_+ + S_-} + \frac{1}{(n-1) \pi} \sum_{c \in \pm} \tan S_c + \order((n-1)^{-3}),
\end{equation}
one finds
\beq
\label{eq:d_s_n}
 d^s_n \defeq (\omega^s_n c_n^2)^2 = \frac{\pi (n-1)}{(S_+ + S_-)^2} + \order((n-1)^{-3}).
\eeq

The parametrix was already computed in the previous section. Taking into account the change in the range of $\sigma$ \wrt the treatment of the massless case, we have, \cf \eqref{eq:ParametrixEnergyScalarMassless},
\[
 \frac{1}{2} ( \del_0 \del_0' + \del_1 \del_1') h_s = - \frac{1}{2 \pi (t+i \eps)^2} + \frac{1}{12 \pi} \frac{1}{\cos^2 \sigma} - \frac{1}{24 \pi} + \order(t).
\]
Using
\beq
\label{eq:SumIdentity}
 \sum_{n=1}^\infty n e^{i ( n +\frac{b}{n}) (t+i \eps)} = - \frac{1}{(t+i\eps)^2} - \frac{1}{12} - b + \order(t), 
\eeq
we may thus write
\begin{multline*}
 \frac{1}{2} ( \del_0 \del_0' + \del_1 \del_1') (w_s - h_s) 
 = \frac{1}{2} d^s_1 e^{i \omega^s_1 t} \\
 + \frac{1}{2} \sum_{n=1}^\infty \left[ d^s_{n+1} e^{i \omega^s_{n+1} (t + i \eps)} - \frac{\pi n}{(S_+ + S_-)^2} e^{i ( \frac{\pi n}{S_+ + S_-} + \frac{\sum \tan S_c}{\pi n}) (t + i \eps)} \right] \\
 + \frac{1}{24 \pi} - \frac{\pi}{24 (S_+ + S_-)^2} - \frac{1}{2 \pi (S_+ + S_-)} \sum_c \tan S_c - \frac{1}{12 \pi \cos^2 \sigma} + \order(t).
\end{multline*}
From \eqref{eq:omega_s_asymptotic}, \eqref{eq:d_s_n} it follows that the sum on the \rhs can be bounded uniformly in $t$ and $\eps$. Furthermore, the resulting local bulk energy density is finite. However, one already sees that this is no longer the case in the massless limit $S_\pm \to \frac{\pi}{2}$. Performing the limit of coinciding points and the integration over $\sigma$, we thus obtain the bulk contribution to the expectation value of the scalar Hamiltonian:
\begin{multline}
\label{eq:H_s_bk}
 \VEV{H^0_{s,\bk}} = \frac{S_+ + S_-}{2} \left( d^s_1 + \sum_{n=1}^\infty \left[ d^s_{n+1} - \frac{\pi n}{(S_+ + S_-)^2} \right] \right) \\ + \frac{S_+ + S_-}{24 \pi} - \frac{\pi}{24 (S_+ + S_-)} - \frac{7}{12 \pi} \sum_{c \in \pm} \tan S_c.
\end{multline}

For the boundary part, we can not use a 1-dimensional Hadamard parametrix, as the boundary field is not a solution to a free wave equation, \cf the source term on the \rhs of \eqref{eq:bc_s}. The boundary quantum field is in fact a generalized free field \cite{Wentzell}. For its renormalization we thus take the following approach: We determine the local singularities and construct a corresponding counterterm out of the local geometric data. Using \eqref{eq:omega_s_asymptotic}, one straightforwardly obtains
\beq
\label{eq:f_s_asymptotic}
 \betrag{f_{s, n}(\pm S_\pm)}^2 = \frac{(S_+ + S_-)^2 \tan^2 S_\pm}{\pi^3 (n-1)^3} + \order((n-1)^{-5})
\eeq
for the normalized mode solutions.
For the two-point function on the boundary, we thus obtain
\begin{align}
% \label{eq:w_bd}
 w^\bd_{s,\pm}(\tau; \tau') & = \sum_{n \in \N_1} \betrag{f_{s, n}(\pm S_\pm)}^2 e^{i \omega^s_n (t + i \eps)} \nn \\
 & = i \frac{(S_+ + S_-) \tan^2 S_\pm}{6} t  +  \frac{\tan^2 S_\pm}{\pi} t^2 \left( \zeta(3) - \frac{3}{4} + \frac{1}{4} \log \frac{- \pi^2 (t + i \eps)^2}{(S_+ + S_-)^2}  \right) + \order(t^3), \nn
\end{align}
where $t = \tau' - \tau$. For the corresponding parametrix, we write distances in terms of the local geometric data, i.e., in terms of
\[
 \rho = \tfrac{1}{2} t^2 R^2 \cos^2 S_\pm,
\]
\cf \eqref{eq:metric}, so that a suitable parametrix is
\[
 h^\bd_{s,\pm} = 
 \frac{\tan^2 S_\pm}{2 \pi R^2 \cos^2 S_\pm} \rho \log \frac{- \rho_\eps}{\Lambda_\pm^2}  + \order(t^3).
\]
Here $\Lambda_\pm$ are renormalization length scales which may depend on the boundary component, at least if the masses at the two endpoints are distinguishable.

For the renormalization of the boundary contribution to the scalar Hamiltonian, we thus have to consider
\begin{align*}
 \del_0 \del'_0 h^\bd_{s,\pm} & = - \frac{\tan^2 S_\pm}{2 \pi} \log \frac{(t+i \eps)^2 R^2 \cos^2 S_\pm}{\Lambda_\pm^2} + \order(t) \\
 & = \sum_{n = 1}^\infty \frac{\tan^2 S_\pm}{\pi n} e^{i \frac{\pi n}{S_+ + S_-} (t + i \eps)} - \frac{\tan^2 S_\pm}{2 \pi} \log \frac{(S_+ + S_-)^2 R^2 \cos^2 S_\pm}{\pi^2 \Lambda^2_\pm} + \order(t).
\end{align*}
The subtraction of the boundary divergences then yields
\begin{align*}
 \VEV{H^0_{s, \bd, \pm}} & = \lim_{t \to 0} \frac{1}{2 \tan S_\pm} \del_0 \del'_0 (w^\bd_\pm - h^\bd_\pm) \\
  & = \frac{1}{2 \tan S_\pm} \left[ e^s_{1,\pm} + \sum_{n = 1}^\infty  \left( e^s_{n+1,\pm} - \frac{\tan^2 S_\pm}{\pi n} \right) \right. \\
  & \qquad \qquad \qquad \left. \vphantom{\sum_{n = 1}^\infty} + \frac{\tan^2 S_\pm}{2 \pi} \log \frac{(S_+ + S_-)^2 R^2 \cos^2 S_\pm}{\pi^2 \Lambda^2_\pm} \right],
\end{align*}
where we used the abbreviation
\[
 e^s_{n, \pm} \defeq (\omega^s_n)^2 \betrag{f_{s, n}(\pm S_\pm)}^2
\]
and the expansions \eqref{eq:omega_s_asymptotic} and \eqref{eq:f_s_asymptotic}.

For the full expectation value of the free scalar Hamiltonian, we thus obtain
\begin{multline*}
 \VEV{H^0_s} = \frac{S_+ + S_-}{2} d^s_1 + \sum_{c \in \pm} \frac{e^s_{1, c}}{2 \tan S_c}  \\
 + \sum_{n=1}^\infty \left( \frac{S_+ + S_-}{2} d^s_{n+1} + \sum_{c \in \pm} \frac{e^s_{n+1, c}}{2 \tan S_c}  - \frac{\pi n}{2(S_+ + S_-)} - \sum_{c \in \pm} \frac{\tan S_c}{2 \pi n} \right) \\
 + \frac{S_+ + S_-}{24 \pi} - \frac{\pi}{24 (S_+ + S_-)} + \sum_{c \in \pm} \frac{\tan S_\pm}{4 \pi} \log \frac{(S_+ + S_-)^2 R^2 \cos^2 S_\pm}{\pi^2 \Lambda^2_\pm},
\end{multline*}
where we absorbed the last term in \eqref{eq:H_s_bk} in a change of the scales $\Lambda_\pm$. With integration by parts, and using the equation of motion \eqref{eq:eom_s}, the boundary condition \eqref{eq:bc_s}, and the normalization condition \eqref{eq:Normalization_s}, one finds
\[
 (S_+ + S_-) d^s_n + \sum_{c \in \pm} \frac{1}{\tan S_c} e^s_{n, c} = \omega^s_n,
\]
so that we may write the above as
\begin{multline}
\label{eq:H_s_final}
 \VEV{H^0_s} = \frac{1}{2} \left[ \omega^s_1 + \sum_{n=1}^\infty \left( \omega^s_{n+1} - \left( \frac{\pi n}{S_+ + S_-} + \sum_{c \in \pm}\frac{\tan S_c}{n \pi} \right) \right) \right] \\
 + \frac{S_+ + S_-}{24 \pi} - \frac{\pi}{24 (S_+ + S_-)} + \sum_{c \in \pm} \frac{\tan S_c}{2 \pi} \log \frac{(S_+ + S_-) R \cos S_c}{\Lambda_c}.
\end{multline}
In particular, only knowledge of the mode frequencies $\omega^s_n$ is required. This expression can thus be seen as the appropriate regularization of \eqref{eq:H_s_formal}.

Let us discuss the renormalization ambiguities in our derivation. For this, it is advantageous to write the scalar part of the action in the proper geometric form
\[
 \mathcal{S}^s_0 = - \frac{1}{2} \int_\Sigma \del_\mu f_s \del^\mu f_s \sqrt{\betrag{g}} \ud^2 x - \frac{1}{2} \sum_{c \in \pm} \frac{R \cos S_c}{\tan S_c} \int_{\del_c \Sigma} \del_a f_s \del^a f_s \sqrt{\betrag{h}} \ud x.
\]
In the second term, the latin indices refer to coordinates on the boundary and are raised with $h^{-1}$. 
The factor $\frac{R \cos S_c}{\tan S_c}$ in front of the boundary term should be seen as a coupling constant. Multiplication of a free action with a constant is compensated by the multiplication of the two-point function with the inverse of that constant. It follows that a factor of $\frac{\tan S_c}{R \cos S_c}$ in front of the boundary parametrix is due to this coupling constant. Let us thus consider the corrected boundary parametrix
\[
 \tilde h^\bd_{s,\pm} = \frac{R \cos S_\pm}{\tan S_\pm} h^\bd_{s,\pm} = - \frac{\kappa_\pm}{2 \pi} \rho \log \frac{- \rho_\eps}{\Lambda_\pm^2}  + \order(t^3),
\]
where we used the geodesic curvature $\kappa_c$, \cf \eqref{eq:GeodesicCurvature}. Hence, this parametrix is constructed out of the local geometric data and changing the scale $\Lambda_c$ amounts to adding a geodesic curvature counterterm at the boundary component $c$.
On the other hand, it is clear that $\tilde h^\bd_{s, \pm} \mapsto \tilde h^\bd_{s, \pm} + \lambda_\pm \kappa \rho$ with some coefficients $\lambda_\pm$ is the only locally constructed redefinition of $\tilde h^\bd_{s, \pm}$ with the correct scaling behavior that contributes to the Hamiltonian. 
In the previous section, we saw that there are no bulk renormalization ambiguities for the scalar part.
So we have seen that the only renormalization ambiguity for the scalar Hamiltonian amounts to the redefinition
\beq
\label{eq:RenormalizationAmbiguity}
 \VEV{H^0_s} \to \VEV{H^0_s} + \sum_{c \in \pm} \lambda_c \tan S_c,
\eeq
corresponding to a geodesic curvature counterterm $\kappa_c \sqrt{\betrag{h_c}}$. Note that, by \eqref{eq:S_Condition}, $\tan S_c \sim \sqrt{\gamma R / m_c} \sim L^{\frac{1}{4}}$ for large $R$, so this is consistent with the fact that an inclusion of geodesic curvature counterterms modifies the Regge trajectory at $\order(L^{\frac{1}{4}})$, \cf \eqref{eq:EH_Classical}.

Let us note that the final result \eqref{eq:H_s_final} could also have been obtained by a point-split regularization of the formal expression \eqref{eq:H_s_formal},
\[
 \VEV{H^0_s(t)} = \frac{1}{2} \sum_{n \geq 1} \omega^s_n e^{i \omega^s_n (t + i \eps)}
\]
combined with a subtraction of the integral over $\sigma$ of the point-split bulk parametrix and the point-split boundary parametrix. We did not take that approach here, as it is a priori not clear whether the integration over $\sigma$ commutes with the limit $t \to 0$. For simplicity, we will perform the calculation of the planar contribution in this way. A calculation analogous to the one performed in the scalar case can be found in Appendix~\ref{app:planar_massive}.

Analogously to the scalar contribution, the formal expression for the expectation value of the planar contribution is
\[
 \langle H^0_q \rangle = \frac{1}{2} \sum_{n \geq 2} \omega^q_n.
\]
The point-split version of this is
\[
 \langle H^0_q(t) \rangle = \frac{1}{2} \sum_{n \geq 2} \omega^q_n e^{i \omega^q_n (t + i \eps)}.
\]
In order to get a finite result, we should subtract the singularity obtained by integration over the contribution from the parametrix. In the coordinates chosen in the present section, the planar parametrix fulfills, \cf \eqref{eq:h_p},
\begin{multline*}
%\label{eq:H_q_bulk_t}
 \tfrac{1}{2} \left( \del_0 \del_0' + \del_1 \del_1' + \tfrac{2}{\cos^2 \sigma} \right) h_p \\
 = - \frac{1}{2 \pi} \left[ \frac{1}{(t+i \eps)^2} + \frac{1}{2 \cos^2 \sigma} \log \frac{- (t+i \eps)^2 R^2 \cos^2 \sigma}{\Lambda^2} + \frac{1}{3 \cos^2 \sigma} + \frac{1}{12} \right] + \order(t).
\end{multline*}
Integration over $\sigma$ yields the following result for the coinciding point divergence due to the bulk:
\begin{multline*}
 - \frac{S_+ + S_-}{2 \pi} \frac{1}{(t+i \eps)^2} - \frac{1}{2\pi} \log \frac{(t+i \eps) R}{\Lambda} \sum_{c \in \pm} \tan S_c \\
 - \frac{S_+ + S_-}{24 \pi} + \frac{S_+ + S_-}{2 \pi} - \frac{1}{2\pi} \sum_{c \in \pm} \tan S_c \log (e \cos S_c) - \frac{1}{6 \pi} \sum_{c\in \pm} \tan S_c.
\end{multline*}
In particular, the bulk contributes a logarithmic divergence, contrary to the scalar sector. However, its coefficient is $\sum_{c \in \pm} \tan S_c$, so that the renormalization ambiguity due to the bulk (by changing the renormalization scale $\Lambda$) is contained in the renormalization ambiguity \eqref{eq:RenormalizationAmbiguity} already determined.

Let us now focus on the boundary contribution. For large $n$, we have the asymptotic behavior
\begin{align}
\label{eq:omega_q_asymptotic}
 \omega^q_n & = \frac{(n-2) \pi}{S_+ + S_-} + \frac{2}{(n-2) \pi} \sum_{c \in \pm} \tan S_c + \order((n-2)^{-3}), \\
 \betrag{f_{p,n}(\pm S_\pm)}^2 & = \frac{(S_+ + S_-)^2 \tan^2 S_\pm}{\pi^3 (n-2)^3} + \order((n-2)^{-5}), \nn \\
 \betrag{f_{r,n}(\pm S_\pm)}^2 & = \frac{4 (S_+ + S_-)^4 \tan^2 S_\pm}{\pi^5 (n-2)^5 \cos^2 S_\pm} + \order((n-2)^{-7}). \nn
\end{align}
Hence, considering \eqref{eq:H_0}, we expect the following logarithmic singularity in the coinciding point limit at the boundary:
\[
 - \frac{1}{2} \sum_{c \in \pm} \frac{\tan S_c}{\pi} \log \frac{\pi (t+i\eps)}{S_+ + S_-}.
\]
As for the scalar contribution, one argues that this divergence should be cancelled by the addition of the counterterm
\[
 \sum_{c \in \pm} \frac{\tan S_c}{2 \pi} \log \frac{(t+i \eps) R \cos S_c}{\Lambda_c}.
\]
However, as for the scalar contribution, it is advantageous to perform the subtraction in the sum, in such a way that the limit of coinciding points can be commuted with the summation limit. One thus obtains
\begin{multline}
\label{eq:H_q_final}
 \VEV{H^0_q} = \frac{1}{2} \left[ \omega^q_2 + \sum_{n=1}^\infty \left( \omega^q_{n+2} - \frac{n \pi}{S_+ + S_-} - \frac{2}{n \pi} \sum_{c \in \pm} \tan S_c \right) \right] \\
  + \frac{S_+ + S_-}{24 \pi} - \frac{\pi}{S_+ + S_-} \frac{1}{24} - \frac{S_+ + S_-}{2 \pi} + \sum_{c \in \pm} \frac{\tan S_c}{\pi} \log \frac{R (S_+ + S_-) \cos S_c}{\Lambda_c},
\end{multline}
where once again we absorbed constant multiples of $\tan S_c$ in a redefinition of $\Lambda_c$.

Our attempts to analytically evaluate \eqref{eq:H_s_final} and \eqref{eq:H_q_final} failed, so that we resort to numerical calculations. For that,
we confine ourselves to the case of identical masses at the endpoints, so in particular $S_+ = S_- = S$ and $\Lambda_+ = \Lambda_- = \Lambda$. The numerical calculation of \eqref{eq:H_s_final} and \eqref{eq:H_q_final} then proceeds as follows:
\begin{enumerate}
\item First, we choose $\frac{\gamma}{m} = 1$ and a grid of values of $R$ and determine the frequencies $\omega^s_n$, $\omega^q_n$ for $n \leq 1000$ and all values of $R$ by taking the general solutions \eqref{eq:ModesGeneral_s}, \eqref{eq:ModesGeneral_p} (with either $B=0$ or $A = 0$) and looking for zeros of the boundary condition. 
One can confirm the asymptotic behavior given by \eqref{eq:omega_s_asymptotic} and \eqref{eq:omega_q_asymptotic}.

\item Due to the asymptotic behavior \eqref{eq:omega_s_asymptotic} and \eqref{eq:omega_q_asymptotic}, the errors due to a cut-off of the sums in \eqref{eq:H_s_final} and \eqref{eq:H_q_final} at some fixed $N$ are asymptotically of $\order(N^{-2})$, with an $R$ dependent coefficient. To correct this, we proceed as follows: We choose a grid in $N$ and determine the expressions \eqref{eq:H_s_final} and \eqref{eq:H_q_final} for the different values of $R$, with the sum cut off at $N$. For fixed $R$, we fit the result with an $c_0 + c_1 N^{-2}$ ansatz in the range $N \in [500,1000]$. The number $c_0$ then gives the result for this $R$.

\item The resulting function of $R$ is then fitted to
\beq
\label{eq:FittingForm}
 C_0 \tan S + C_1 + C_2 R^{-\frac{1}{2}}
\eeq
in the range $R \in [100,1000]$.
The first term corresponds to the renormalization ambiguity and is thus not relevant. The second term, however, directly yields the intercept (up to the sign), according to the discussion below \eqref{eq:E2Intercept}.
\end{enumerate}

Note that the contribution of the last term in \eqref{eq:H_s_final} and \eqref{eq:H_q_final} to the target space energy behaves asymptotically as $R^{-\frac{1}{2}} \log R$, i.e., it slightly dominates the renormalization ambiguity. The quality of the fits to \eqref{eq:FittingForm} indicates\footnote{One can also include such a term into the fits and finds that it has a very small coefficient.} that it has been properly subtracted, yielding a test of our renormalization prescription.

Our method yields the values
\begin{align*}
 C_1^s & \simeq - 0.00001 &
 C_1^q & \simeq - 1.00001
\end{align*}
for the scalar and the planar part. These results are quite robust under changes of the fitting range or the fitting function. We interpret them as being the numerical approximation of
\begin{align*}
 C_1^s & = 0 &
 C_1^q & = - 1,
\end{align*}
corresponding to the intercept \eqref{eq:a1}.

Let us comment on the implications of the result for the Nambu-Goto string as a phenomenological model for hadrons. For measured meson trajectories and the endpoint masses and the intercepts as free parameters, intercepts in the range $a \in [-0.55, 0]$ were found \cite{SonnenscheinWeissman14} (for a fit to the orbital angular momentum), in plain contradiction with the theoretical value $a = 1$. However, one has to keep in mind that our semi-classical calculation is only valid for large angular momenta. The maximum spin which was used to determine the trajectories in \cite{SonnenscheinWeissman14} was $L = 6$. But $6^{\frac{1}{4}} \simeq 1.57$, so $L^{\frac{1}{4}}$, $L^0$ and $L^{-\frac{1}{4}}$ are all of the same order. It seems doubtful that one can consistently distinguish between these contributions with so little data. Apart from that, the model is of course rather crude in that it neglects, for example, the spin of the quarks. However, it is conceivable that fixing $a$ to the theoretical value yields a more consistent assignment of quark masses and the $\alpha$ parameter of the Einstein-Hilbert term \eqref{eq:S_EH} to the different trajectories.

\section{Degeneracies of excited states}
\label{sec:Degeneracy}

For the semi-classical spectrum of excitations of the massless open string with a fixed angular momentum component $L_{1,2}$, we found oscillators with frequency $n \geq 1$ for each of the $D-3$ directions perpendicular to the plane of rotation and oscillators with frequency $n \geq 2$ for excitations in the plane of rotation. The goal of this section is to compare with the spectrum of excitations of the covariantly quantized open string, i.e., to investigate the physical states that are eigenstates of the energy and of angular momentum $L_{1,2} = \ell$. A particular focus will be on the presence or absence of an $n=1$ excitation in the plane of rotation.

In the covariantly quantized Nambu-Goto string, the state of minimal energy for a fixed angular momentum $\ell$ in the $1-2$ plane is given by (for simplicity, we fix $\pi \gamma = 1$)
\[
 \ket{\ell} = (\xi \cdot \alpha_{-1})^\ell \ket{0,\sqrt{2 ( \ell - a)}}.
\]
Here
\[
 \xi = \tfrac{1}{\sqrt{2}} (0, 1, i, 0, \dots, 0) 
\]
and $\ket{0, m}$ stands for the ground state with vanishing spatial momentum and rest mass $m$. We recall the definitions, \cf \cite{GreenSchwarzWittenI},
\begin{align*}
 L_m & = \frac{1}{2} \sum_{n = - \infty}^\infty \alpha_{m-n} \cdot \alpha_n, & m \neq 0 \\
 L_0 & = \frac{1}{2} \alpha_0^2 + \sum_{n =1}^\infty \alpha_{-n} \cdot \alpha_n,  \\
 J^{\mu \nu} & = - i \sum_{n = 1}^\infty \frac{1}{n} \left( \alpha^\mu_{-n} \alpha^\nu_n - \alpha^\nu_{-n} \alpha^\mu_n \right),
\end{align*}
and the commutation relations
\beq
 \label{eq:alphaCommutator}
 [\alpha^\mu_m, \alpha^\nu_n] = m \delta_{m + n} \eta^{\mu \nu}.
\eeq
In order to avoid confusion with the Virasoro generators $L_m$, we here switch to the notation $J^{\mu \nu}$ for the angular momentum. We also omitted the center-of-mass contribution to $J^{\mu\nu}$. 
We note that $\alpha_0 = p$, the momentum operator. The commutation relations \eqref{eq:alphaCommutator} imply
\begin{align*}
 [L_m, \zeta \cdot \alpha_{-k}] & = k \zeta \cdot \alpha_{m-k}, \\
 [J^{1 2}, \zeta \cdot \alpha_{-k}] & = \tilde \zeta \cdot \alpha_{-k},
\end{align*}
where
\[
 \tilde \zeta = (0, - i \zeta^2, i \zeta^1, 0, \dots, 0).
\]
With the last equation, one straightforwardly checks that $\ket{\ell}$ is an eigenstate of $J^{12}$ of eigenvalue $\ell$. Furthermore, one checks that the state $\ket{\ell}$ is physical, i.e., it fulfills the conditions
\begin{align*}
 \left( L_m - \delta_{0 m} a \right) \ket{\ell} & = 0 & \forall m \geq 0.
\end{align*}
Finally, \eqref{eq:alphaCommutator} implies that $\ket{\ell}$ has positive norm.

Let us begin by considering the minimal excitations of $\ket{\ell}$, i.e., the physical states which are eigenstates of $J^{12}$ with eigenvalue $\ell$ and of $p^2$ with eigenvalue $2 ( \ell + 1 - a)$. It is easy to find $D-3$ linearly independent states:
\[
 \zeta \cdot \alpha_{-1} (\xi \cdot \alpha_{-1})^\ell \ket{0,\sqrt{2 ( \ell + 1 - a)}}.
\]
Here $\zeta$ is an element of the subspace spanned by $e_3$ -- $e_{D-1}$. These correspond to the $D-3$ scalar excitations for $n=1$ of the semi-classical open rotating string. These states obviously have positive norm, so they count as proper physical excitations.

We can see the $D-3$ linearly independent operators $\zeta \cdot \alpha_{-1}$ as the creation operators for the oscillator of frequency $n=1$. As a slight complication, also the momentum needs to be shifted and when powers of these operators are applied, correction terms need to be added to ensure physicality. For example,
\[
 \left[ (\zeta \cdot \alpha_{-1})^2 - \tfrac{\zeta^2}{1- 2 p^2} \left( \tfrac{1}{p^2} (p \cdot \alpha_{-1})^2 - p \cdot \alpha_{-2} \right) \right] (\xi \cdot \alpha_{-1})^\ell \ket{0, \sqrt{2 (\ell + 2 - a)}}
\]
is the state obtained by twice acting with $\zeta \cdot \alpha_{-1}$ and adding corrections to ensure physicality. Similarly, one may see the $D-3$ linearly independent operators $\zeta \cdot \alpha_{-n}$ as the creation operators for the oscillator with frequency $n$, up to correction terms.
In the scalar sector, we thus have complete agreement of the spectra of the semi-classical and the covariantly quantized Nambu-Goto string.

The analog of the first excitation of the planar $n=2$ mode is given by
\[
 \left[ \bar \xi \cdot \alpha_{-1} \xi \cdot \alpha_{-1} - \tfrac{\ell + 1}{1- 2 p^2} \left( \tfrac{1}{p^2} (p \cdot \alpha_{-1})^2 - p \cdot \alpha_{-2} \right) \right] (\xi \cdot \alpha_{-1})^\ell \ket{0, \sqrt{2 (\ell + 2 - a)}}.
\]
This state has positive norm, at least in the range $a \leq 2$. Higher excitations of this mode are constructed by applying $\bar \xi \cdot \alpha_{-1} \xi \cdot \alpha_{-1}$ several times, and adding correction terms to ensure physicality. Similarly, excitations of the $n$th planar mode are obtained by acting with $\bar \xi \cdot \alpha_{-n+1} \xi \cdot \alpha_{-1}$ and applying correction terms. This exhausts the excitation spectrum of the semi-classical string.

However, there is also a state corresponding to a planar $n=1$ mode:
\[
 \left[ \xi \cdot \alpha_{-2} ( \xi \cdot \alpha_{-1} )^{\ell-1} - 2 p^{-2} p \cdot \alpha_{-1} ( \xi \cdot \alpha_{-1} )^{\ell} \right] \ket{0,\sqrt{2 ( \ell + 1 - a)}}.
\]
It is straightforward to check that this is an eigenstate of $J^{12}$ of eigenvalue $\ell$ and of $L_0$ with eigenvalue $a$. Also the physicality conditions are fulfilled.
 However, one finds that this state has positive norm for $a < 1$, is null for $a = 1$, and has negative norm for $a > 1$. In the critical covariantly quantized string, i.e., with $a = 1$, this state would thus not correspond to a physical excitation. In this sense, the spectra of excitations in the semi-classical and the critical covariantly quantized string coincide.

\subsection*{Acknowledgements}

I would like to thank M.~Wrochna for helpful discussions on spectral calculus in Krein spaces, S.~Hellerman and I.~Swanson for clarifying discussion on the relation to their work \cite{HellermanSwanson}, and M.~Kozon for discussions on locally covariant renormalization that were helpful for finding the mistake in a previous version of this manusript.

\appendix

\section{The boundary conditions}
\label{app:bc}

The boundary is a submanifold of co-dimension $D-1$, so in addition to the scalar and planar perturbations, also radial perturbations could be relevant there. To the \rhs of \eqref{eq:phi_Parametrization}, we thus add $f_r v_r$ with $v_r = (0, \cos \tau, \sin \tau, 0)$.

To work out the implication of the boundary condition \eqref{eq:bc} on the perturbations $\vp$, we first determine the variation of the metric (the brackets denote symmetrization in $\mu$, $\nu$):
\begin{align*}
 \delta g_{\mu \nu} & = 2 \del_{(\mu} \bar X_a \del_{\nu)} \vp^a  \\
 & = 2 \del_{(\mu} \bar X_a \del_{\nu)} v_p^a f_p + 2 \del_{(\mu} \bar X_a \del_{\nu)} v_r^a f_r + 2 \del_{(\mu} \bar X_a v_r^a \del_{\nu)} f_r  \\
 & = 2 R \left( \left[ f_p - \tfrac{1}{2} \sin \sigma \dot f_r \right] \begin{pmatrix} 0 & 1 \\ 1 & 0 \end{pmatrix} + \cos \sigma f_r \begin{pmatrix} 1 & 0 \\ 0 & 0 \end{pmatrix} - \sin \sigma f_r' \begin{pmatrix} 0 & 0 \\ 0 & 1 \end{pmatrix} \right).
\end{align*}
Here we used that the vectors $v_s$, $v_p$ are orthogonal to the world-sheet, that $\del_\nu v_s = 0$ and
\begin{align*}
 \del_0 \bar X & = R \cos \sigma \sin \sigma v_p + R \sin^2 \sigma e_0, \\
 \del_1 \bar X & = - R \sin \sigma v_r, \\
 v_p' & = - \cot \sigma v_p - e_0,
\end{align*}
with $e_0$ the unit vector in time direction. This implies
\begin{align*}
 \delta \sqrt{\betrag{g}} & = - R \cos \sigma f_r - R \sin \sigma f_r', \\
 \delta g^{\mu \nu} & = \frac{2 f_p - \sin \sigma \dot f_r}{R^3 \sin^4 \sigma} \begin{pmatrix} 0 & 1 \\ 1 & 0 \end{pmatrix} - \frac{2 \cos \sigma f_r}{R^3 \sin^4 \sigma} \begin{pmatrix} 1 & 0 \\ 0 & 0 \end{pmatrix} + \frac{2 f_r'}{R^3 \sin^3 \sigma} \begin{pmatrix} 0 & 0 \\ 0 & 1 \end{pmatrix}.
\end{align*}
We thus obtain
\[
 \delta \left[ \sqrt{\betrag{g}} g^{1 \nu} \del_\nu X \right] = \cot \sigma f_r v_r + f_s' v_s + \left( \cot \sigma f_p - \cos \sigma \dot f_r + f_p' \right) v_p
 + \left( f_p - \sin \sigma \dot f_r \right) e_0.
\]
Linear independence of $v_p, v_s, v_r, e_0$ implies that $f_r = f_p = f_s' = 0$ at the boundary. Furthermore, with l'H\^opital's rule, we also obtain $f_p' = 0$.

\section{The planar two-point function}
\label{app:2pt}

To compute the \lhs of \eqref{eq:2pt_p_eval}, we have to evaluate
\[
 \sum_{n = 2}^\infty \frac{1}{4 n} \left[ \left( n^2 + \frac{2}{\sin^2 \sigma} \right) f_{p, n}^2 + { f'_{p,n} }^2 \right] e^{i n (t + i \eps)}
\]
Straightforward manipulations simplify this to
\begin{multline*}
 \sum_{n = 2}^\infty \frac{n}{2 \pi (n^2 -1)} \left[ n^2 + \cot^2 \sigma + \frac{2}{\sin^2 \sigma} \cos 2 n \sigma - \frac{3 \cot \sigma}{n \sin^2 \sigma} \sin 2 n \sigma \right. \\
 \left. + \frac{2 \cos^2 \sigma + 1}{n^2 \sin^4 \sigma} \sin^2 n \sigma \right] e^{i n (t + i \eps)}.
\end{multline*}
Using
\begin{align*}
 \sum_{n = 2}^\infty \frac{n^3}{n^2 -1} e^{i n (t + i \eps)} & = - \frac{1}{(t+i\eps)^2} - \frac{1}{2} \log [- (t + i \eps)^2] - \frac{11}{6} + \order(t), \\
 \sum_{n = 2}^\infty \frac{n}{n^2 -1} e^{i n (t + i \eps)} & = - \frac{1}{2} \log [- (t + i \eps)^2] - \frac{3}{4} + \order(t), \\
 \sum_{n = 2}^\infty \frac{n \cos 2 n \sigma}{n^2 -1} e^{i n (t + i \eps)} & = - \frac{1}{2} - \frac{1}{4} \cos 2 \sigma - \frac{1}{2} \cos 2 \sigma \log [4 \sin^2 \sigma] + \order(t), \\
 \sum_{n = 2}^\infty \frac{\sin 2 n \sigma}{n^2 -1} e^{i n (t + i \eps)} & = \frac{1}{4} \sin 2 \sigma - \frac{1}{2} \sin 2 \sigma \log [4 \sin^2 \sigma] + \order(t), \\
 \sum_{n = 2}^\infty \frac{\sin^2 n \sigma}{n(n^2 -1)} e^{i n (t + i \eps)} & = \frac{3}{4} \sin^2 \sigma - \frac{1}{2} \sin^2 \sigma \log [4 \sin^2 \sigma] + \order(t),
\end{align*}
one obtains the \rhs of \eqref{eq:2pt_p_eval}.

\section{The calculation of the planar part}
\label{app:planar_massive}

In this appendix, we want to discuss the calculation of the planar part in the massive case in the same fashion as for the scalar part, i.e., without assuming that integration over $\sigma$ and the limit $t \to 0$ commute. For simplicity, we assume equal masses, i.e., $S_+ = S_- = S$.

Let us first concentrate on the bulk. For odd (even) $n$, the (anti-) symmetric planar mode $f_{p, n}$ is realized. With the normalization given in \eqref{eq:ModesGeneral_p} with $A (B) = 1$, we obtain (we set $f_n = f_{p, n}$, $\omega_n = \omega^q_n$)
\begin{multline}
\label{eq:PlanarModeEnergy}
 \left( \omega_n^2 + \tfrac{2}{\cos^2 \sigma} \right) f_{n}(\sigma)^2 + f'_{n}(\sigma)^2 = \omega_n^4 + \omega_n^2 \tan^2 \sigma \pm \omega_n^2 \tfrac{2}{\cos^2 \sigma} \cos 2 \omega_n \sigma \\ \pm 3 \omega_n \tfrac{\tan \sigma}{\cos^2 \sigma} \sin 2 \omega_n \sigma + \tfrac{2 \sin^2 \sigma + 1}{2 \cos^4 \sigma} ( 1 \mp \cos 2 \omega_n \sigma).
\end{multline}
Asymptotically, the normalization constants $c_n$, that have to be multiplied to $f_n$ for the normalization \eqref{eq:Normalization}, fulfill
\[
 c_n^2 \omega_n^4 = \frac{\pi}{4 S^2} (n-2) + \frac{1}{\pi (n-2)} + \order((n-2)^{-3}),
\]
so that only the first three terms on the \rhs of \eqref{eq:PlanarModeEnergy} contribute to the $t \to 0$ singularity of the two-point function. We denote their sum by $T_n(\sigma)$. The remaining terms are denoted by $R_n(\sigma)$. Using \eqref{eq:SumIdentity} and
\begin{align*}
 \sum_{n = 1}^\infty \frac{1}{n} e^{i n (t + i \eps)} & = - \frac{1}{2} \log - (t + i \eps)^2 + \order(t), \\
 \sum_{n = 1}^\infty \frac{(-1)^n}{n} e^{i n (t + i \eps)} \cos ( 2 n \sigma) & = - \frac{1}{2} \log 4 \cos^2 \sigma + \order(t) & \betrag{\sigma} < \tfrac{\pi}{2},
\end{align*}
we may thus write
\begin{multline*}
 \tfrac{1}{2} \left( \del_0 \del_0' + \del_1 \del_1' + \tfrac{2}{\cos^2 \sigma} \right) (w_q - h_q) = \tfrac{1}{2} c_2^2 T_2(\sigma) \\
 + \frac{1}{2} \sum_{n=1}^\infty \left[ \vphantom{\frac{1}{\pi n \cos^2 \sigma}} c_{n+2}^2 T_{n+2}(\sigma) e^{i \omega_{n+2} (t + i \eps)} - \frac{\pi n}{4 S^2} e^{i (\frac{\pi n}{2 S} + \frac{4 \tan S}{\pi n}) (t+i \eps)} \right. \\
   \left. - \frac{1}{\pi n \cos^2 \sigma} \left( 1 - (-1)^n 2 \cos 2 \tfrac{\pi n}{2 S} \sigma \right) e^{i \frac{\pi n}{2 S} (t + i \eps)} \right] \\
   + \frac{1}{2} \sum_{n = 2}^\infty c_n^2 R_n(\sigma) + \frac{1}{24 \pi \cos^2 \sigma} - \frac{\pi}{96 S^2} - \frac{\tan S}{\pi S} + \frac{1}{24 \pi} \\
   + \frac{1}{4 \pi \cos^2 \sigma} \log \frac{4 S^2 R^2 \cos^2 \sigma}{\pi^2 \Lambda^2 } + \frac{1}{2 \pi \cos^2 \sigma} \log 4 \cos^2 \tfrac{\pi}{2S} \sigma.
\end{multline*}
In this expression, we may take the limit $t \to 0$, to obtain the bulk energy density
\begin{multline*}
 \tfrac{1}{2} \left( \del_0 \del_0' + \del_1 \del_1' + \tfrac{2}{\cos^2 \sigma} \right) (w_q - h_q) = \tfrac{1}{2} c_2^2 T_2(\sigma) \\
 + \frac{1}{2} \sum_{n=1}^\infty \left[ c_{n+2}^2 T_{n+2}(\sigma) - \frac{\pi n}{4 S^2} - \frac{1}{\pi n \cos^2 \sigma} \left( 1 - (-1)^n 2 \cos 2 \tfrac{\pi n}{2 S} \sigma \right) \right] \\
   + \frac{1}{2} \sum_{n = 2}^\infty c_n^2 R_n(\sigma) + \frac{1}{24 \pi \cos^2 \sigma} - \frac{\pi}{96 S^2} - \frac{\tan S}{\pi S} + \frac{1}{24 \pi} \\
   + \frac{1}{4 \pi \cos^2 \sigma} \log \frac{4 S^2 R^2 \cos^2 \sigma}{\pi^2 \Lambda^2 } + \frac{1}{2 \pi \cos^2 \sigma} \log 4 \cos^2 \tfrac{\pi}{2S} \sigma.
\end{multline*}
Due to \eqref{eq:PlanarModeEnergy} and the asymptotic forms of $c_n$ and $\omega_n$, the sum can be dominated uniformly in $\sigma$ for $\sigma \in [-S, S]$ and $S < \frac{\pi}{2}$. Hence, except for the last term, the energy density is bounded for $\sigma \in [-S, S]$ for $S < \frac{\pi}{2}$. The logarithmic divergence of the energy density near the boundary is a well-known phenomenon in two-dimensional massive scalar field theories, \cf \cite{GribMamayevMostepanenko} for example. We have thus established that the energy density is integrable and integration over $\sigma$ yields
\begin{multline*}
 \VEV{H^0_{q, \bk}} = \tfrac{1}{2} \int c_2^2 (T_2(\sigma) + R_2(\sigma)) \ud \sigma \\
 + \frac{1}{2} \sum_{n=1}^\infty \left[ \int c_{n+2}^2 (T_{n+2}(\sigma) + R_{n+2}(\sigma) ) \ud \sigma - \frac{\pi n}{2 S} - \frac{2 \tan S}{\pi n} + Q_n \right] \\
   + \frac{\tan S}{12 \pi} - \frac{\pi}{48 S} - \frac{2 \tan S}{\pi} + \frac{S}{12 \pi}
   -\frac{S}{\pi}  + \frac{\tan S}{\pi} + \frac{\tan S}{2 \pi} \log \frac{4 S^2 R^2 \cos^2 \sigma}{\pi^2 \Lambda^2 } + Q,
\end{multline*}
with
\begin{align*}
 Q_n & = (-1)^n \int_{-S}^S \frac{2}{\pi n \cos^2 \sigma} \cos \tfrac{\pi n}{S} \sigma \ud \sigma, \\
 Q & = \int_{-S}^S \frac{1}{2 \pi \cos^2 \sigma} \log 4 \cos^2 \tfrac{\pi}{2S} \sigma \ud \sigma.
\end{align*}
Using integration by parts, one shows that $\betrag{Q_n} < C n^{-2}$ and
\[
 \frac{1}{2} \sum_{n=1}^\infty Q_n = - Q,
\]
so that the above reduces to
\begin{multline*}
 \VEV{H^0_{q, \bk}} = \frac{1}{2} \int c_2^2 (T_2(\sigma) + R_2(\sigma)) \ud \sigma \\
 + \frac{1}{2} \sum_{n=1}^\infty \left[ \int c_{n+2}^2 (T_{n+2}(\sigma) + R_{n+2}(\sigma) ) \ud \sigma - \frac{\pi n}{2 S} - \frac{2 \tan S}{\pi n} \right] \\
   - \frac{\pi}{48 S} + \frac{S}{12 \pi} -\frac{S}{\pi}  + \frac{\tan S}{2 \pi} \log \frac{4 S^2 R^2 \cos^2 \sigma}{\pi^2 \Lambda^2 },
\end{multline*}
where we absorbed terms of the form $C \tan S$ in a change of the scale $\Lambda$.

For the boundary component, one obtains, analogously to the scalar part,
\[
 \VEV{H^0_{q, \bd}} = \frac{1}{\tan S} \left[ e^q_2 + \sum_{n = 1}^\infty \left( e^q_{n+2} - \frac{\tan^2 S}{\pi n} \right) + \frac{\tan^2 S}{2 \pi} \log \frac{4 S^2 R^2 \cos^2 S}{\pi^2 \Lambda^2} \right]
\]
where we used
\[
 e^q_n \defeq c_n^2 \left[ \left( \omega_n^2 - \tfrac{1}{\cos^2 S} \right) f_{p, n}(S)^2 + \left( \omega_n^2 - 1 - 2 \tan^2 S \right) \betrag{f_{r, n}}^2 \right].
\]
In total, we thus have
\begin{multline*}
 \VEV{H^0_{q}} = \frac{1}{2} \int c_2^2 (T_2(\sigma) + R_2(\sigma)) \ud \sigma + \frac{1}{\tan S} e^q_2 \\
 + \frac{1}{2} \sum_{n=1}^\infty \left[ \int c_{n+2}^2 (T_{n+2}(\sigma) + R_{n+2}(\sigma) ) \ud \sigma + \frac{2}{\tan S} e^q_{n+2} - \frac{\pi n}{2 S} - \frac{4 \tan S}{\pi n} \right] \\
   - \frac{\pi}{48 S} + \frac{S}{12 \pi} -\frac{S}{\pi}  + \frac{\tan S}{\pi} \log \frac{4 S^2 R^2 \cos^2 \sigma}{\pi^2 \Lambda^2 }.
\end{multline*}
Using integration by parts, the equations of motion \eqref{eq:eom_p}, \eqref{eq:bc_p}, \eqref{eq:bc_r}, and the normalization \eqref{eq:Normalization_p}, one finds
\[
 \frac{1}{2} \int c_n^2 (T_n(\sigma) + R_n(\sigma)) \ud \sigma + \frac{1}{\tan S} e^q_n = \frac{1}{2} \omega^q_n.
\]
Hence, we obtain \eqref{eq:H_q_final} for the special case $S_+ = S_- = S$.

%\bibliography{../mybib}{}
%%%\bibliographystyle{../halpha}
%%\bibliographystyle{../h-elsevier}
%\bibliographystyle{../h-elsevier_new}

\end{document}